\documentclass[aps, prd, amsmath, floats, floatfix, twocolumn,
superscriptaddress, nofootinbib, showpacs]{revtex4-1}
\usepackage[xetex]{graphicx}
\usepackage{color}
\usepackage{soul}
\usepackage{url}
\usepackage{bm}         % bold math symbols
\usepackage{times}

\newcommand{\beq}{\begin{equation}}
\newcommand{\eeq}{\end{equation}}
\newcommand{\beqn}{\begin{eqnarray}}
\newcommand{\eeqn}{\end{eqnarray}}

\newcommand{\lo}{\mathrel{\raise.3ex\hbox{$<$}\mkern-14mu
    \lower0.6ex\hbox{$\sim$}}}
\newcommand{\go}{\mathrel{\raise.3ex\hbox{$>$}\mkern-14mu
    \lower0.6ex\hbox{$\sim$}}}

\newcommand{\RNum}[1]{\uppercase\expandafter{\romannumeral #1\relax}}

\usepackage{color}

\newcommand{\Cornell}{\affiliation{Center for Radiophysics and Space
    Research, Cornell University, Ithaca, New York, 14853, USA}}
\newcommand{\WSU}{\affiliation{Department of Physics \& Astronomy,
	Washington State University, Pullman, Washington 99164, USA}}
\newcommand{\CITA}{\affiliation{Canadian Institute for Theoretical 
    Astrophysics, University of Toronto, Toronto, Ontario M5S 3H8, Canada}}

\newcommand{\CIFAR}{\affiliation{Canadian Institute for Advanced Research, 180 Dundas St.~West, Toronto, ON M5G 1Z8, Canada}}
\newcommand{\TAPIR}{\affiliation{TAPIR, Walter Burke Institute for Theoretical Physics,
    California Institute of Technology, MC 350-17, Pasadena, CA 91125, USA}}
\newcommand{\YITP}{\affiliation{Center for Gravitational Physics and International Research Unit of Advanced Future Studies,
    Yukawa Institute for Theoretical Physics, Kyoto University, Kyoto, Kyoto Prefecture 606-8317, Japan}}
\newcommand{\IUCAA}{\affiliation{Inter-University Centre for Astronomy and Astrophysics (IUCAA), 
    Post Bag 4, Ganeshkhind, Pune 411 007, India}} %
\newcommand{\BERKELEY}{\affiliation{Lawrence Berkeley National Laboratory, 1 Cyclotron Rd, Berkeley, CA 94720, USA}} %
\newcommand{\NEWHAMPSHIRE}{\affiliation{Department of Physics, University of New Hampshire, Durham, New Hampshire 03824, USA}} %
\newcommand{\JINA}{\affiliation{Joint Institute for Nuclear Astrophysics, Michigan State University, East Lansing, MI 48824, USA}} % 
\newcommand{\NCSU}{\affiliation{Department of Physics, North Carolina State University, Raleigh, NC 27695, USA}} % 
\newcommand{\NCSA}{\affiliation{NCSA, University of Illinois at Urbana-Champaign, Urbana, Illinois, 61801, USA}} %
\newcommand{\AEI}{\affiliation{Max Planck Institute for Gravitational Physics (Albert Einstein Institute), Am M\"{u}hlenberg 1, Potsdam-Golm, 14476, Germany}} % 

\usepackage{graphicx}% Include figure files
\usepackage{dcolumn}% Align table columns on decimal point
\usepackage{bm}% bold math
\usepackage{epsf}

\begin{document}

\title{Evolution of the Magnetized, Neutrino-Cooled Accretion Disk in the 
  Aftermath of a Black Hole  Neutron Star Binary Merger}

\author{Fatemeh Hossein Nouri} \WSU\IUCAA %
\author{Matthew D. Duez} \WSU % 
\author{Francois Foucart} \BERKELEY\NEWHAMPSHIRE %

\author{M. Brett Deaton} \JINA\NCSU\WSU %
\author{Roland Haas} \NCSA\AEI\TAPIR %
\author{Milad Haddadi} \WSU %
\author{Lawrence E. Kidder} \Cornell %
\author{Christian D. Ott} \TAPIR\YITP %
\author{Harald P. Pfeiffer} \CITA\CIFAR\AEI %
\author{Mark A. Scheel} \TAPIR %
\author{Bela Szilagyi} \TAPIR %

\begin{abstract}
Black hole-torus systems from compact binary mergers are possible engines for gamma-ray bursts (GRBs). 
During the early evolution of the post-merger remnant, the state of the torus is determined by a combination of neutrino
cooling and magnetically-driven heating processes, so realistic models must include both effects. 
In this paper, we study the post-merger evolution of a magnetized black hole-neutron star binary system
using the Spectral Einstein Code (SpEC) from an initial post-merger state provided by previous
numerical relativity simulations.  We use a finite-temperature nuclear equation of state and incorporate
neutrino effects in a leakage approximation.  To achieve the needed accuracy, we introduce improvements
to SpEC's implementation of general-relativistic magnetohydrodynamics (MHD), 
including the use of cubed-sphere multipatch grids and an improved method
for dealing with supersonic accretion flows where primitive variable recovery is difficult.
We find that a seed magnetic field triggers a sustained
source of heating, but its thermal effects are largely cancelled
by the accretion and spreading of the torus from MHD-related angular momentum transport. The neutrino
luminosity peaks at the start of the simulation, and then 
drops significantly over the first 20\,ms but in roughly the same way for magnetized and
nonmagnetized disks.  The heating rate and disk's luminosity decrease much more slowly thereafter.  These
features of the evolution are insensitive to grid structure and resolution, formulation of the MHD
equations, and seed field strength, although turbulent effects are not fully converged.  
%Evolving system up to 60ms after merger for our 
%strongly-cooled disk and chosen seed field, we do not see the formation
%of strong outflows.

\end{abstract}


\maketitle

\section{Introduction}
\label{sec:intro}

%This paper is a continuation of~\cite{Foucart:2014nda}.

%Introduction outline:

%- short GRBs:
The cause of short-hard gamma ray bursts (GRBs) remains unknown,
but some of the most promising central engine models involve
rapid ($\sim M_{\odot}$ s${}^{-1}$) accretion onto a stellar mass
black hole (BH). Such systems are naturally produced by some black
hole-neutron star (BHNS) and neutron star-neutron star (NSNS)
binary mergers.  (For reviews of short GRBs, see~\cite{Nakar2007,Berger:2013jza}.)

Given the requisite dense, hot accretion flow, there are several ways
% NDAFs model require high accretion rate to explain GRBs (Narayan et al 2001) 
energy could be channeled into a baryon-poor ultra-relativistic
outflow of the sort needed to explain GRB properties.  The accretion gas
cools primarily by neutrino emission, and so such systems are
classified as neutrino-dominated accretion flows
(NDAFs)~\cite{Popham99,2002ApJ...579..706D,2004MNRAS.355..950J}.  Some
emitted neutrino energy can be transferred to a pair fireball through
neutrino-antineutrino annihilations outside the
disk~\cite{1991AcA....41..257P,Jaroszynski:1995mf,richers:15,Just:2015dba}.
Magnetic fields can also extract energy from the disk or black hole
spin~\cite{1977MNRAS.179..433B,Meszaros:1996ww}, and the energy outflow can be Poynting flux dominated.

The lifetime of a short GRB ($\lesssim 1 \mathrm{s}$, presumably related to the disk
lifetime $\tau_{\rm acc}$) is much greater than
the dynamical timescale ($\tau_{\rm d} \sim$ ms) and perhaps also the thermal
timescale ($\tau_{\rm th}\sim \alpha^{-1}\tau_{\rm d} \sim (H/r)^2\tau_{\rm acc}$
in the standard alpha viscosity, thin disk model~\cite{shakura:1973}).  Therefore,
the GRB mechanism is a process that takes place in the
accretion system's dynamical and probably also thermal equilibrium.

The post-merger accretion disks formed in BHNS/NSNS mergers have
densities of $\rho \sim 10^{11}\,\mathrm{g\,cm}^{-3}$ and
temperatures of $T\sim 1\,\mathrm{MeV}$. Hence, photons are trapped
and in equilibrium, and radiation is by neutrinos.
For high enough accretion
rate $\dot{M}$, the disk is opaque to neutrinos, which must diffuse
out and provide an additional source of pressure.  Neutrino
luminosities can reach $L_{\nu} \sim 10^{53}-10^{54}$ erg s${}^{-1}$, and this
emission will strongly affect the disk (on a secular timescale
$\tau_{\rm th}$) by cooling it and altering the composition, quantified
by the electron fraction $Y_e$, the fraction of nucleons that are protons.
Unstable entropy or $Y_e$ gradients can drive convection in the
disk~\cite{Lee:2005se}.  In addition to these emission effects, there are also
neutrino transport effects.  Neutrino absorption near the neutrinosphere
can drive thermal winds~\cite{Dessart2009,Perego2014};
neutrino momentum transport can create
a viscosity that slows the growth of the magnetorotational
instability~\cite{Masada2008,Guilet2014} (although probably not
%once the 
%hypermassive merger remnant collapses
%to a BH and also not
for BHNS mergers~\cite{Foucart:2015vpa,Kiuchi:2015qua}).  

In previous papers~\cite{Deaton2013,Foucart:2014nda,Foucart:2015vpa},
we simulated BHNS mergers at realistic mass ratios using a
finite-temperature nuclear equation of state and incorporating
neutrino effects.  The latter were modeled in some cases with a
leakage scheme (which includes emission but not
transport)~\cite{Ruffert1996,Rosswog:2003rv,OConnor2010,Deaton2013,Foucart:2014nda}
and in some cases with an energy-integrated two-moment M1 transport
scheme~\cite{shibata:11,Foucart:2015vpa}.  
Comparing to the earlier times of evolution we found that the post-merger
accretion disks become cold, and more neutron-rich with dimmer neutrino
% after long evolution (40ms) see fig 14 and 17 from [19]
emission after a few tens of milliseconds.  Comparing leakage to M1,
we find that the former gives a reasonable estimate for the neutrino emission and
global thermal evolution, although it overestimates temperature
gradients, and cannot accurately track the $Y_e$ evolution in
low-density regions. %, and also gives the total $Y_e$ much lower (by $\sim0.05$).  
No significant neutrino-driven winds were seen.
The cooling and dimming of the disks is unsurprising, given that these
simulations included the major cooling mechanisms--neutrino emission
and advection of the hot inner gas into the black hole--but contained
only one significant heating mechanism (in addition to numerical dissipation heating):
shock heating from the circularization and pulsation of the disk gas.

Long-term accretion requires an angular momentum transport process
that will naturally release orbital energy and heat the gas.  This is
probably provided by magnetic fields, which were not included in the
above simulations.  Weakly magnetized accretion flows are subject to
the magnetorotational instability (MRI)~\cite{1998RvMP...70....1B},
inducing turbulence which dissipates energy at small scales and whose
mean (mostly Maxwell) stresses transport angular momentum outward,
driving accretion~\cite{Hawley:1995sy}.  Magnetic fields also transport angular
momentum through magnetic winding (the $\omega$ effect).  Reconnection
at current sheets provides a way to convert magnetic energy into
plasma thermal and kinetic energy.  Simulations of radiatively
inefficient magnetized accretion tori find strong winds along disk
surfaces and magnetically dominated
poles~\cite{DeVilliers:2003gy,mckinney:04}.  Large-scale fields
threading the BH ergosphere enables extraction of the black hole
spin energy into a Poynting flux-dominated
jet~\cite{1977MNRAS.179..433B,mckinney:04}.

There have been successful GRMHD simulations, neglecting neutrino
effects, of
BHNS~\cite{Chawla:2010sw,2012PhRvD..85f4029E,2013arXiv1303.0837E,Paschalidis2014,Kiuchi:2015qua}
and
NSNS~\cite{Liu:2008xy,Anderson:2008zp,Giacomazzo:2010bx,Rezzolla:2011da,Kiuchi2014,Kiuchi2015}
mergers.  The highest resolution BHNS simulations with an initial seed
field confined in the neutron star~\cite{Kiuchi:2015qua} find strong
winds and Poynting-dominated jets only at very high resolutions (and
even here, it is unclear that convergence has been achieved).
There are also indications that unconfined seed fields produce jets
more readily~\cite{Paschalidis2014}, consistent with disk studies that
find jets but not disk interiors to be very sensitive to the seed
field~\cite{Beckwith:2007sr}.  The helicity of the magnetic field may
also have subtle long-term effects~\cite{Wan:2016yid}.  These merger
simulations used simplified thermal components of the equation of
state and neglected neutrino effects; they had the main heating
effects but not a major cooling effect.

Clearly, accurate evolution on thermal timescales requires both
neutrino cooling and magnetoturbulent heating.  The two will influence
each other.  The neutrino luminosity, and hence the viability of
``neutrino'' mechanisms for driving a GRB, depends on magnetic
heating, while the saturation strength of the magnetic field in an MRI
turbulent disk will depend on the temperature of the
gas~\cite{Sano:2003bf,Shi:2015mvh} set partly by neutrino cooling.
NSNS merger simulations with both effects have been
performed~\cite{Neilsen:2014hha,Palenzuela2015}, but our understanding
of long-term post-merger evolution of BHNS (and high-mass NSNS)
systems relies on accretion disk models.  In most cases, turbulent
transport and dissipation is modeled by a phenomenological ``alpha''
viscosity~\cite{shakura:1973}.  These include the original
one-dimensional (axisymmetric, vertically summed), equilibrium NDAF
studies~\cite{Popham99,2002ApJ...579..706D,2004MNRAS.355..950J}.
One-dimensional NDAFs were evolved by Janiuk~{\it et al.}~\cite{janiuk:07}, 
who found disks can become visco-thermally
unstable in some regions, but only for very high accretion rates
($\dot{M} \gtrsim 10\,M_\odot\,\mathrm{s}^{-1}$).  Evolutions have also been carried out
in higher dimensions, again in the alpha viscosity framework, yielding
valuable information on neutrino-antineutrino energy release and
late-time
outflows~\cite{Setiawan:2004,Lee:2005se,Fernandez2013,Fernandez:2014}.
Efficient release of energy by radiation requires low $\alpha$ (so
$\tau_{\rm acc}>\tau_{\rm th}$), proving~\cite{Lee:2005se} the importance of
% \tau_{\rm acc} \sim (H/R)^(-2)
first-principles, magnetohydrodynamic (MHD)
simulations to assess the adequacy of viscosity
models and to reveal the actual efficiency of angular momentum
transport.

MHD disk simulations with neutrino cooling have been carried out in 2D
beginning from analytic, constant angular momentum equilibrium tori
by several
groups~\cite{2007PThPh.118..257S,2011NewA...16...46B,Shibata:2012zz,Janiuk:2013lna}
, and recently in 3D by Siegel and Metzger~\cite{Siegel2017}.
They identify the MRI, with associated heating, neutrino emission,
and powerful outflows.  These studies probably provide the most realistic
picture available of the evolution of the post-merger disk, but their artificial
disk profiles neglect the strong angular momentum gradients, high compactness,
and nonaxisymmetric features seen in merger simulations.  These neglected
features will most likely have strong effects in the early, and most
neutrino luminous, post-merger phase.  In addition, the 2D (axisymmetric)
simulations~\cite{2007PThPh.118..257S,2011NewA...16...46B,Shibata:2012zz,Janiuk:2013lna}
are affected by the known
differences between the saturation of the MRI in 2D vs. 3D~\cite{Goodman:1994,obergaulinger:09},
including the impossibility of an axisymmetric dynamo~\cite{Hawley:1992,1996ApJ...465L.115B}.

%- About our simulation
In this paper, we study the effects of magnetic fields 
on the post-merger evolution of a BHNS binary system.  We evolve in 3D
using as initial data the BH accretion flow produced by
a BHNS merger simulation~\cite{Foucart:2014nda}.  
In addition to MHD, we employ a realistic finite-temperature nuclear equation
of state and neutrino cooling via a leakage approximation, giving us all the
basic ingredients needed for a realistic thermal evolution.  For this first study,
we restrict ourselves to a simple seed field geometry with high field strength,
for which the MRI is resolved with modest grid sizes.  Studying a strongly magnetized disk
most likely gives us a sense of the maximum effect that magnetic fields can
have.  Our simulations use the Spectral Einstein Code (SpEC) and required
the development of new numerical techniques for SpEC:  MHD on a cubed-sphere
multipatch grid, coordinate maps to optimize grid use, and an improved technique
to control entropy evolution in regions where kinetic energy dominates over
internal energy.

Comparing disk evolutions with and without magnetic fields, we find
some expected effects.  The magnetic field drives strong and sustained
accretion, while the late-time accretion rate of a nonmagnetized disk
is, by comparison, negligible. Magnetic effects also do increase the
disk's specific entropy, as a result of magnetoturbulent heating  
 and numerical reconnection, leading to a roughly steady entropy in
comparison to the secularly decreasing entropy of a nonmagnetized
disk.  However, at early times the nonmagnetized disk's cooling rate is significantly
slower than neutrino emission would predict, indicating the continued
importance of shock heating 30\,ms after merger as a heating source of
comparable strength to MHD-related heating.  The effects of disk
depletion and heating on the neutrino luminosity roughly cancel, and
the magnetized disk dims at roughly the same rate as the
nonmagnetized disk.  Thus, for the case we consider,
MHD turbulence does little to assist
neutrino-related mechanisms for powering a GRB during the most neutrino
luminous phase of the accretion, even in the case of an
extremely strong seed field.

This paper is organized as follows.  In Sec.\RNum{2}, the initial
configuration and set up is discussed.  Section \RNum{3} briefly describes
the numerical methods used.  In Sec.\RNum{4}, numerical results are
presented, focusing on the effects of magnetic field on the  accretion
rate, thermal evolution and general properties of the disk.  Finally,
Sec.\RNum{5} is devoted to the summary and conclusion.  A detailed
discussion of new numerical techniques is reserved for the Appendix.

\section{Initial state}
\label{sec:methods}

\subsection{Input physics}
As in our recent BHNS merger studies~\cite{Deaton2013,Foucart:2014nda} we employ
the Lattimer-Swesty equation of state~\cite{Lattimer:1991nc} with nuclear 
incompressibility $K_0 = 220\,\mathrm{MeV}$ (LS220), using the table available at 
http://www.stellarcollapse.org and described in ~\cite{OConnor2010}. 

Neutrino emission effects are captured using a simple leakage scheme,
described in~\cite{Deaton2013,Foucart:2014nda}.  Leakage schemes
remove  energy and alter lepton number at rates based on the local
free-emission and diffusion rates.  They account for these emission
effects within factors of $\sim 2-3$ accuracy (as determined by comparisons with
genuine neutrino transport schemes~\cite{Foucart:2015vpa}) but do not include the
effects of neutrino transport and absorption.  Our leakage scheme
integrates out spectral information, assuming Fermi-Dirac
distributions at the local temperature (with chemical potentials
estimated as in~\cite{Foucart:2014nda}), although we can estimate an
average energy of emitted neutrinos from the total luminosity and
number emission rate. (See~\cite{Perego2014,Perego:2016} for approximate ways, not pursued
in this study, to incorporate absorption and spectral information
in a leakage framework.) Our leakage scheme includes $\beta$-capture
processes, $e^+-e^-$ pair annihilation,  plasmon decay and
nucleon-nucleon Bremsstrahlung interactions.  In optically thick
regions, the neutrinos contribute to the pressure.

\subsection{Initial Configuration}
For our initial state, we use the BHNS configuration  M12-7-S9 presented
in~\cite{Foucart:2014nda}. (See Table 2 of that paper.)  The initial
masses of the BH and NS are $7 M_{\odot}$ and $1.2 M_{\odot}$
respectively.  The BH is rapidly spinning with $S_{\rm
  BH}/M_{\rm BH}{}^2 = 0.9$. The remnant torus mass is about  $0.14
M_{\odot}$, with maximum density of $\sim 2\times 10^{12}$ g
cm${}^{-3}$,  and average temperature of $\sim 2.7$ MeV. We restart
our simulation using data of this case at $t = 15\,\mathrm{ms}$ after merger.  At
this time, the spacetime has settled to a nearly stationary BH
metric in the coordinate system produced by the numerical
relativity simulation, but the disk remains significantly
nonaxisymmetric and nonstationary.  We therefore evolve only the
fluid, keeping the metric at its initial state.
%We study several cases with different physics input, and using different methods to handle 
% BH singularity and energy evolution. (see Table~\ref{tab:diskprop}).

We set up an initially poloidal magnetic field via a toroidal vector
potential
\begin{equation}
A_\phi = A_b \varpi^2 \max(\rho-\rho_\text{cut}, 0) \,,
\end{equation}
where $\rho$ is the axisymmetrized density field (to initiate the field with large poloidal loops), 
$\varpi = \sqrt{x^2 + y^2}$ is the cylindrical radius in grid
coordinates,  $A_b$ sets the overall strength of the resulting
$B$-field, and the cutoff density $\rho_\text{cut}$, set to 6\% of the maximum density,
confines the initial field to regions of high-density matter.
%We choose the initial field strength to have (for our strong magnetic field case)
%\begin{equation}
%\left(\frac{\int \rho \beta^{-1} d^{3}x}{\int \rho d^{3}x}\right)^{-1} \approx 13,
%\end{equation}
%defining $\beta = P_\mathrm{gas} / P_B$ and $\rho=\rho_0 \sqrt{g} W$. 
We follow the same prescription as that in Noble~{\it et al.}~\cite{Noble:2008tm} 
to set the initial magnetic field strength,
so that the ratio of the volume-weighted integrated gas pressure to 
the volume-weighted integrated magnetic pressure $\equiv \beta$ 
is about 13 for our strongly magnetized disk.
This magnetic field at the maximum value is about
$3.8 \times 10^{15} \,\mathrm{G}$.  This is likely much
stronger than realistic BHNS post-merger magnetic fields.  We focus on
this extreme case first for two reasons.  First, it allows us to
resolve the rapidly-growing modes of the MRI very well with modest
resolution.  Second, an extreme field might be expected to reveal the
maximum effect that magnetic fields might have.

Strong seed fields may induce qualitatively different behavior from weaker
seeds if it is strong enough to suppress the MRI before the disk can
become turbulent.  This will certainly be the case where $\beta$ is near
or below unity, so the fastest-growing
MRI mode wavelength $\lambda_{\rm MRI}$ 
($\sim (2\pi/\Omega)(B/\sqrt{4\pi\rho})$)
exceeds the disk height.  This
is not a danger in most of our disk.  However, MRI growth might also
be affected if $\lambda_{\rm MRI}$ is comparable to the length scale on
which $\lambda_{\rm MRI}$ itself varies (due to variation in
Alfven speed)~\cite{Piontek:2003in,Pino:2008iz,obergaulinger:09}
 or comparable to the radius of
curvature of the field lines, which occurs even in some strong-field, high-density
regions.
 In order to estimate the effect of seed field strength, 
we carry out another simulation with a weaker
seed field, set by $\beta \approx  36$. This corresponds
to a maximum field strength of $2\times 10^{15}$~G.
This simulation does show weaker heating and less outflow,
confirming our expectation (also suppported by 2D strong-seed
disk simulations~\cite{Janiuk:2013lna}) that a strong field maximizes
MHD-related effects.

\begin{table*}
\begin{ruledtabular}
\begin{tabular}{ l l l l l l l }
%\toprule
Name & $N_i{}^a$ & $\Delta r {}^b$(m) & $\Delta z {}^b$(m) & BH singularity${}^c$ & Energy evolution${}^c$ & $<\beta>_{\rm init}$ \\
%\midrule
\toprule 
B0-P-$\tau$-L0 & 213 & 2880 & 580 &  Puncture & $\tau$ & $\infty$ \\
B0-P-$\tau$-L1 & 266 & 2285 & 467 & Puncture & $\tau$ & $\infty$ \\                         
$\beta$13-P-$\tau$-L0 & 213 & 2880 & 580 & Puncture & $\tau$ & 13 \\
$\beta$13-P-$\tau$-L1 & 266 & 2285 & 467 & Puncture & $\tau$ & 13 \\
$\beta$13-P-$\tau$-L2 & 332 & 1806 & 376 & Puncture & $\tau$ & 13 \\
B0-M-$\tau$-L1 & 178 & 2360 & 573 & Multipatch & $\tau$ & $\infty$ \\
$\beta$13-M-$\tau$-L1 & 178 & 2360 & 573 & Multipatch & $\tau$ & 13 \\
B0-M-Ent-L0 & 138 & 3128 & 763 & Multipatch & Entropy & $\infty$ \\
B0-M-Ent-L1 & 178 & 2360 & 573 & Multipatch & Entropy & $\infty$ \\
B0-M-Ent-L1r & 178 & 1675 & 573 & Multipatch & Entropy & $\infty$ \\
$\beta$13-M-Ent-L0 & 138 & 3128 & 763 & Multipatch & Entropy & 13 \\ 
$\beta$13-M-Ent-L1 & 178 & 2360 & 573 & Multipatch & Entropy & 13 \\ 
$\beta$13-M-Ent-L1r & 178 & 1675 & 573 & Multipatch & Entropy & 13 \\ 
$\beta$13-M-Ent-L2 & 231 & 1790 & 426 & Multipatch & Entropy & 13 \\ 
$\beta$36-M-Ent-L1 & 178 & 2360 & 573 & Multipatch & Entropy & 36 \\ 
$\beta$36-M-Ent-L2 & 231 & 1790 & 426 & Multipatch & Entropy & 36 \\
$\beta$36-P-$\tau$-L1 & 266 & 2285 & 467 & Puncture & $\tau$ & 36 \\
%\bottomrule
\end{tabular}
\end{ruledtabular}
\caption{\label{tab:diskprop} A list of simulations reported. 
Simulations vary by grids, numerical methods, and strength of
seed field. \\
${}^a$ The cube of $N_i$ is the total number of grid points. \\
${}^b$  $\Delta r$ and $\Delta z$ are the radial and vertical grid
spacing, respectively, on the equator at the radius of the initial density
maximum. \\
${}^c$ See Appendix for details.
}
\end{table*}

\section{Numerical Methods}

Previous SpEC hydrodynamics simulations evolved fluids on Cartesian grids
with points inside a radius $r_{\rm EX}$ inside the BH horizon excised,
resulting in an irregular-shaped cubic-sphere or ``legosphere'' excision region.  Points
within a stencil of $r_{\rm EX}$ were evolved with one-sided differencing. 
This proved numerically unstable for MHD evolutions--an unsurprising result given
the presence of incoming characteristic speeds on legosphere boundary
faces.

\begin{figure}
  \includegraphics[width=\linewidth]{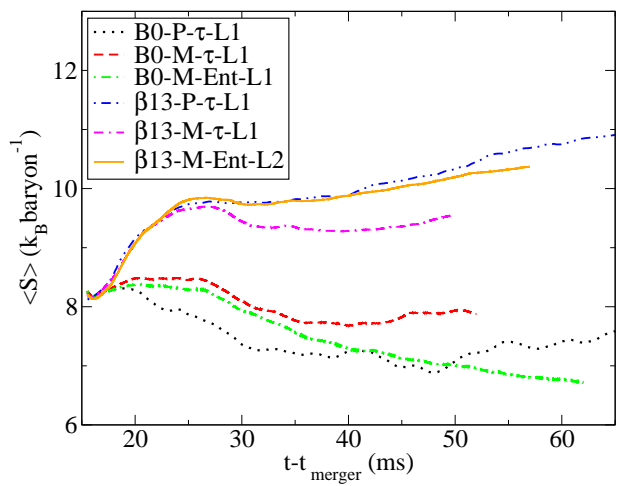}
  \caption[S-Cases-Methods]{ Comparison of the total density-averaged
    entropy for different numerical methods for nonmagnetized and
    stronger magnetized cases.
%~\ref{sec:numerical-methods}.
    The $\tau$ method for energy evolution shows extra heating happening at the late time evolution
    for the nonmagnetized case using both puncture and multipath methods.  
    }
  \label{fig:S-Cases-Methods}
\end{figure}

\begin{figure}
  \includegraphics[width=\linewidth]{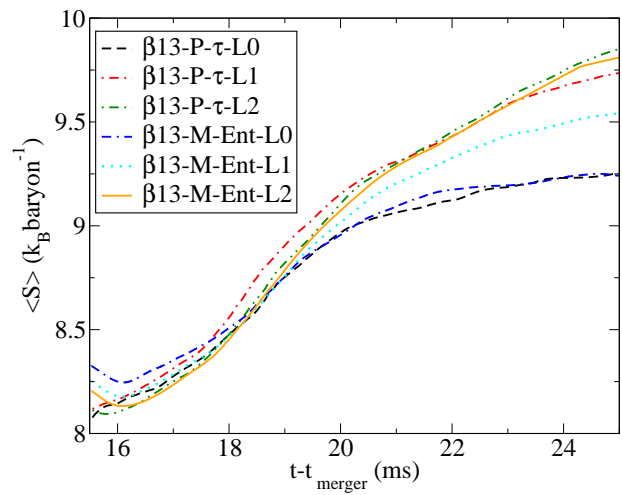}
  \caption[S-ResTest-Methods]{
  Convergence test on the specific entropy in the first $10\,\mathrm{ms}$ of evolution for
  the magnetized disk with puncture-tau and multipatch-entropy methods.  A smoothing of
  scalar primitive variables after the interpolation onto multipatch grids causes slightly
  higher initial average entropy in these runs, but the difference quickly decreases and
  the subsequent evolutions for both methods are in good agreement at different resolutions.  
  }
  \label{fig:S-ResTest-Methods}
\end{figure}

\begin{figure}
  \includegraphics[width=\linewidth]{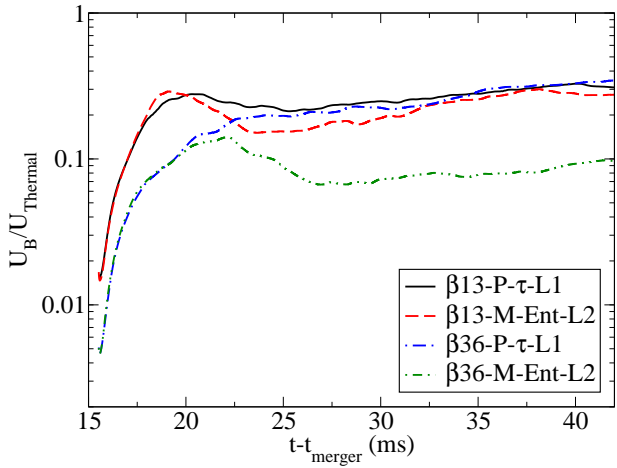}
  \caption[UB-Over-UThermal-Methods-Cases]{
  Comparison of the magnetic energy to the thermal energy ratio for different 
magnetized cases with different methods. The $\beta$36-P-$\tau$-L1 case reaches
the same saturation level as the stronger field cases, but the magnetic energy starts
to dissipate in the $\beta$36-M-Ent-L2 case, leading the ratio to decrease over time
and finally saturate at a lower level 
(see table~\ref{tab:diskprop} for simulation labels).
  }
  \label{fig:UB-Over-UThermal}
\end{figure}

We implemented two fixes to enable stable magnetized inflow into the
BH.  The first is
to map to a new coordinate system in which the sphere $r_{\rm EX}$ is
mapped to a point, so that the interior of this sphere is not on
the grid ("excision by coordinates"). 
This method is implicitly used in non-vacuum numerical relativity
moving puncture evolutions~\cite{Baiotti:2006wm,Faber:2007dv,Liu:2009}
and has been explicitly used for MHD
by Etienne~{\it et~al}~\cite{Etienne:2010ui}. 
We then evolve the MHD equations as in~\cite{Muhlberger2014} with
constrained transport and no explicit excision.  We call this a
``puncture'' method.  The second fix is to replace Cartesian grids
with cubes deformed so as to fit together and fill the space between inner and
outer spherical shells, the so called ``cubed-sphere'' configuration
which has already been successfully applied by other codes to
numerical relativity~\cite{Thornburg:2004dv,Schnetter:2006pg,Reisswig:2006nt,Pazos:2009vb,Pollney:2009yz},
hydrodynamics~\cite{Zink:2007xn,2011PhRvD..83d3007K,korobkin:12,reisswig:13a},
and MHD~\cite{Koldoba:2002kx,Fragile:2008ca,Romanova:2012}. 
Each deformed sphere is evolved on its local coordinate system.
We call this method ``multipatch''.  For
the induction equation, we implement a centered hyperbolic divergence
cleaning method.  Details of these methods and code tests are provided
in the Appendix.

An additional numerical challenge is posed by the nonmagnetized disk
which, as it cools, becomes more supersonic.  In our conservative MHD
formulation, only the total energy and momentum density are evolved,
so it becomes difficult to accurately extract temperature information
when internal energy is much less than kinetic energy.  SpEC has a
procedure~\cite{Muhlberger2014} for ``fixing'' energy and momentum
evolution variables when they fail to map to any physical temperature
and velocity.  In previous papers, this fixing was invoked only in
unimportant low density regions, but here it leads to glitches in
temperature inside the high-density region of the torus and unphysical
heating.  We cure this problem by introducing an auxiliary entropy
variable used to exclude unphysical jumps in temperature, similar
to a technique used in the HARM3D code~\cite{Noble:2008tm}.  Details
are given in the Appendix.

A list of the combinations of methods and resolutions reported in
this paper is provided in Table~\ref{tab:diskprop}.  A comparison
of results for the average entropy evolution is given in
Fig.~\ref{fig:S-Cases-Methods}.  Entropy is a particularly useful
diagnostic of thermal evolution because it responds only to
physical heating and cooling effects.  Unlike temperature, entropy
is unaffected by adiabatic expansion/compression and by nuclear reactions
(if, as here, the gas remains in nuclear statistical equilibrium).
We see that the methods give overall agreement, except that only
simulations with the new entropy variable can maintain cooling
of the nonmagnetized disk.

In Fig.~\ref{fig:S-ResTest-Methods}, we test convergence of
magnetized disk runs by evolving with both grid types at three
resolutions.  Fortunately, puncture and multipatch runs seem to
converge to each other.  Puncture grids have more gridpoints
for a given resolution of the disk interior, but they also
allow larger timesteps (because they don't have the multipatch
code's concentration of angular grid points near the horizon).

For the nonmagnetized disk evolution, we have investigated the effect
of numerical viscosity on the late-time cooling rate.  We evolve in
multipatch mode at three resolutions (the same as in the magnetized
disk convergence test).  We also perform a fourth simulation with a
radial map that concentrates resolution near the maximum-density ring,
increasing resolution there by a factor of 2.5.  (See Appendix for
details.)  In all cases, the
entropy curves, and especially the late-time cooling slopes, are
nearly identical.  We conclude that numerical viscosity cannot be an
important part of the energy budget for this disk's evolution.

In the magnetized disk simulations it is essential to resolve the 
MHD instabilities to capture all MHD effects. 
Resolving the MRI requires high resolution 
($\approx 10$ grid points, to capture the growth of fastest-growing mode, 
along $\lambda_{\rm MRI}$~\cite{2006PhRvD..74j4026S}).  
We achieve this resolution despite a modest number of grid zones
by using strong seed fields and by using coordinate maps to increase
the resolution in high density regions near the disk midplane as
described in the Appendix.
Measuring $\lambda_{\rm MRI}/\Delta x$ at the initial 
time shows that MRI fastest-growing mode is resolvable in over $80\%$ of the magnetized fluid (medium resolution). 
We find that the thermal evolution is much more sensitive to vertical than
to radial resolution, presumably because it is the mode of the axisymmetric
MRI with vertical wavenumber that is most significant in the high-density
region, so we use grids with $\Delta z < \Delta r$.
% HARM3d resolution: (192,192,64) for (r,theta,phi)

Although it is simple to check $\lambda_{\rm MRI}/\Delta x$, it is not possible
to disentangle MRI-driven field amplification from other effects.  Local field
amplification on the orbital time is seen--in fact, it is seen even in some regions
where the MRI fastest-growing mode is certainly not resolved, as would be expected
from nonmodal shearing wave amplification~\cite{Squire:2014cra}.  In our case, there is the
additional complication that our initial state is not a hydrodynamic equilibrium
but an extremely dynamical configuration. 

Fig.~\ref{fig:UB-Over-UThermal} shows 
another comparison of results for the total magnetic energy to the thermal energy
ratio for the magnetized cases. There is a good agreement
between the puncture and multipatch methods for the stronger field case; 
The energy ratio grows by more than
one order of magnitude and saturates at the same level for both methods.
For the weaker field case, the multipatch run does not resolve MRI growth
as well, so there is a larger difference.  For the puncture run, the magnetic field
saturates at the same level as the stronger field, indicating that the saturation state
is independent of the initial seeded magnetic field (at least for our range of seed fields).
The weakly magnetized-multipatch simulation (case $\beta$36-M-Ent-L2 in table~\ref{tab:diskprop})
on the other hand, tracks the similar puncture simulation $\beta$36-P-$\tau$-L1 
for about $5\,\mathrm{ms}$, and then it decreases for about $10\,\mathrm{ms}$
and finally saturates at a level that is lower by a factor of two.
This shows that our puncture method can resolve the magnetic field growth better
for weakly magnetized case.
Based on the methods comparison and convergence studies, we present puncture simulation 
for the weakly magnetized case ($\beta=36$), and multipatch simulations for 
the nonmagnetized and strongly magnetized ($\beta=13$) cases in the next section. 

%Below, when presenting results for magnetized 
%evolutions of global quantities and magnetic fields, we use puncture simulations.
%We use multipatch simulations only for nonmagnetized evolution and radial profiles of matter fields. 
%The reason which makes the magnetic energy dissipate in the 
%multipatch run is still under the inverstigation. Therefore we decided to present
%the results of the puncture simulations for the magnetized cases in the results section.    

%\results
\section{Results}
\label{sec:Results}

We concentrate only on the results of simulations using multipatch grid and auxiliary
entropy evolution methods with moderate resolution for the nonmagnetized case (B0-M-Ent-L1),
and high resolution for strongly magnetized case ($\beta$13-M-Ent-L2),
and the puncture $\tau$ evolution methods for the weakly magnetized case 
($\beta$36-P-$\tau$-L1) in Table~\ref{tab:diskprop}.
%In addition to these simulations, we also use the results from $\beta$13-M-Ent-L2
%for our stronger magnetized case, because they converged better for the matter fields.  
All grids and evolution methods give similar results for the
first $\sim$25\,ms, but these particular runs give more reasonable results in the subsequent
evolution (see the detailed discussion in appendix~\ref{sec:numerical-methods}).  
At the initial time, the thermal timescale is
estimated as $\tau_{\rm thermal}\sim E_{\rm thermal}/L_{\nu}\sim 10$\,ms.  We evolve for about 50\,ms,
long enough to see the disk altered by thermal effects.

% dynamical evolution of the disk
\subsection{Dynamical evolution}

\begin{figure*}
  \includegraphics[width=\linewidth]{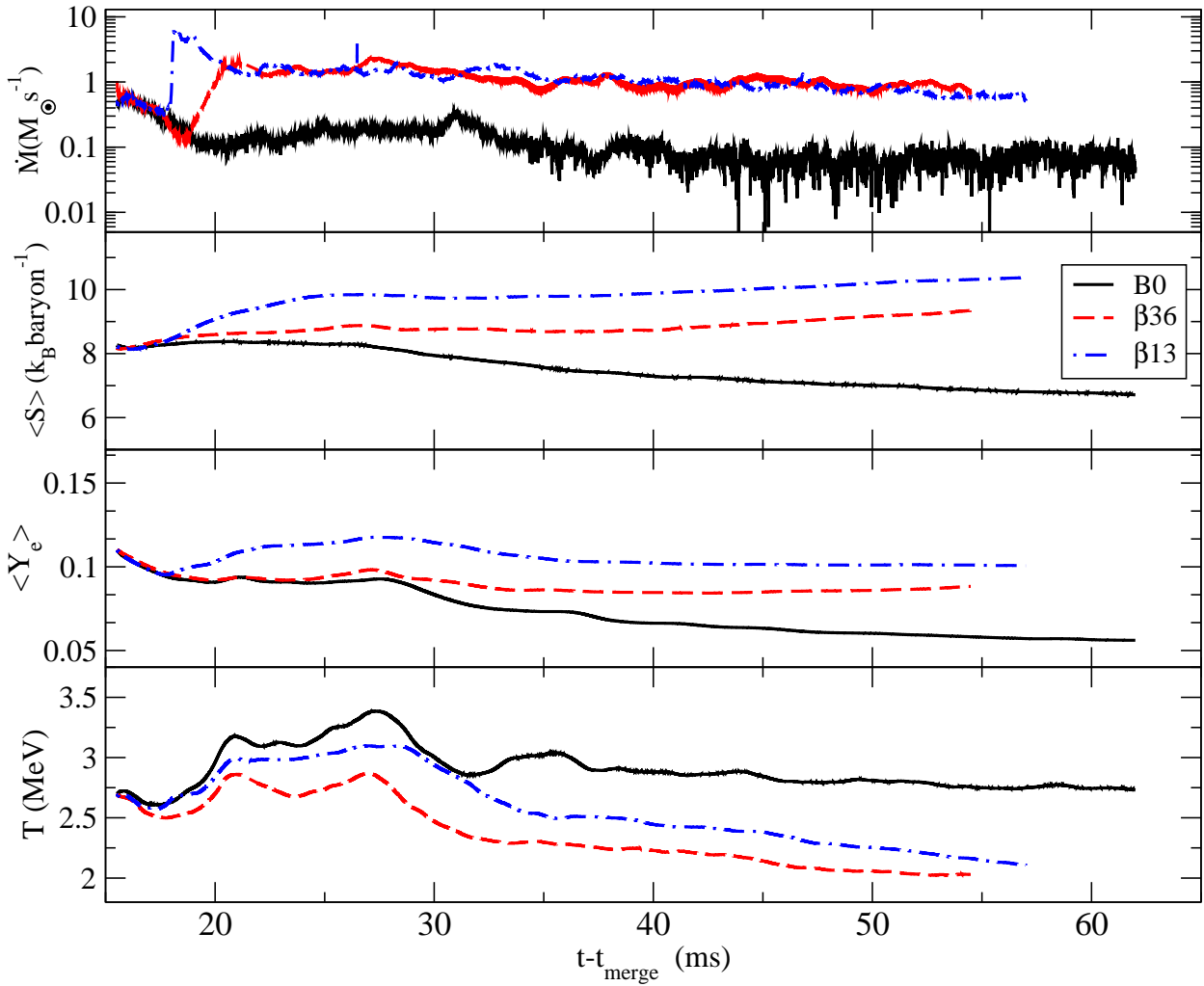}
  \caption[Global-B0-beta13-PTau]{
The evolution of the accretion rate $\dot{M}$ (first panel from the top), 
the electron fraction $Y_e$ (second panel),
the specific entropy $s$ (third panel), and temperature $T$ (last panel) for 
magnetized $\beta$13-P-$\tau$-L1 (dashed dot line), $\beta$36-P-$\tau$-L1 (dashed line), 
and nonmagnetized B0-M-Ent-L1 cases (solid line). 
The accretion rate is higher by about one order of magnitude 
for the magnetized cases due to magnetorotational instablity. 
The entropy grows higher 
as a result of effective viscous heating, while the temperature decreases over time because of
adiabatic cooling for the magnetized cases. 
  }
  \label{fig:Global-B0-beta15-PTau}
\end{figure*}

\begin{figure}
  \includegraphics[width=\linewidth]{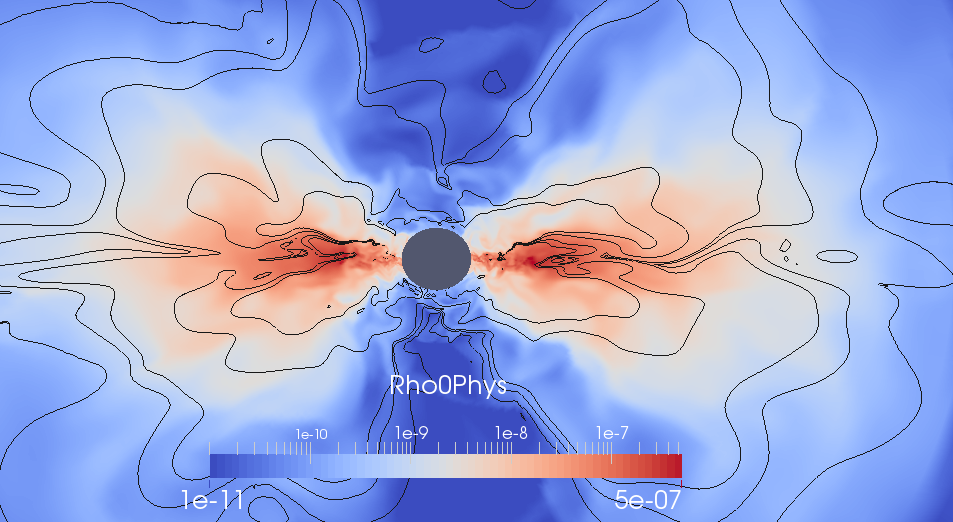}
  \caption[Rho0Phys-B-contour]{
Snapshot of the rest-mass density in the meridional x-z plane
at $t = 45\,\mathrm{ms}$ for $\beta$13 case.
The solid line shows magnetic field magnitude contours
correspond to $\approx$ [$10^{12}$,
$10^{13}$,$10^{14}$,$10^{15}$,$10^{16}$]\,G.
  }
  \label{fig:Rho0Phys-B-contour}
\end{figure}

\begin{figure}
  \includegraphics[width=\linewidth]{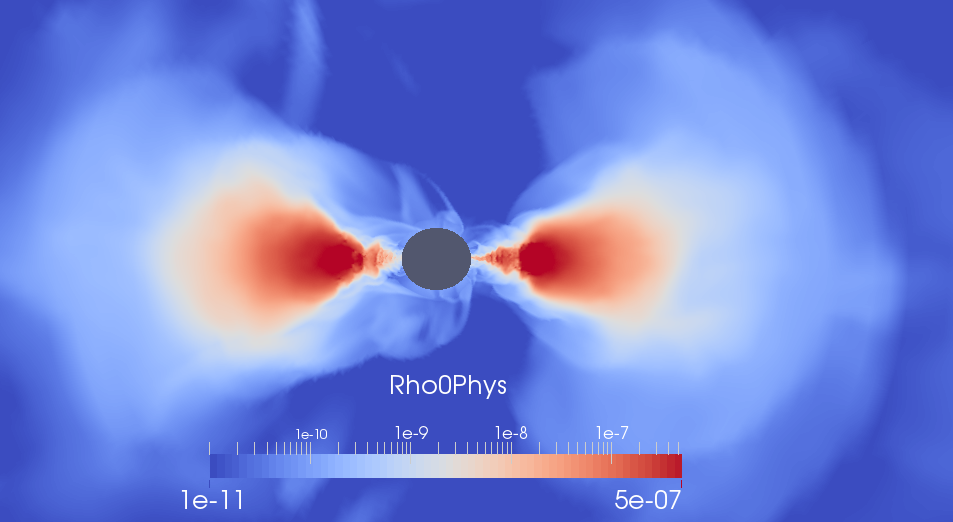}
  \caption[Rho0Phys-B0]{
Snapshot of the rest-mass density in the meridional x-z plane
at $t = 45\,\mathrm{ms}$ for the nonmagnetized case.
  }
  \label{fig:Rho0Phys-B0}
\end{figure}

In Fig.~\ref{fig:Global-B0-beta15-PTau}, we plot several global
quantities of the disk.  As expected, adding a magnetic field enables
angular momentum transport, leading to an accretion rate roughly one
order of magnitude higher than that of the disk evolved without a
magnetic field.  The accretion rate does not appear very sensitive
to the strength of the seed field, at least for the very limited range
studied here.  The settled accretion rate of $\sim
0.4\,M_\odot\,\mathrm{s}^{-1}$ is low enough that a thermal
instability is not expected~\cite{janiuk:07}.  MHD effects can also
cause the disk to expand radially and vertically,  as is expected from
angular momentum transport (see the 2D images of the
density profile Figs.~\ref{fig:Rho0Phys-B0} and ~\ref{fig:Rho0Phys-B-contour}
showing the nonmagnetized and magnetized disks at $t=45$\,ms respectively).  
This transport especially drives matter into the inner radii ($r \sim 30\,\mathrm{km}$), leading
to higher densities there.  The nonmagnetized disk, on the other hand,
contracts vertically and radially, becoming more ring-like as it loses
thermal pressure support. Evolution without a magnetic field leads to
a significantly denser disk, which explains why the magnetized disks
have lower average temperature even though they have higher average
entropy (last panel in Fig.~\ref{fig:Global-B0-beta15-PTau}).

The lower three panels of Fig.~\ref{fig:Global-B0-beta15-PTau}
show the effect of magnetic fields on the average entropy per baryon
$\langle S\rangle$, electron fraction $\langle Y_e\rangle$, and
temperature $\langle T\rangle$.  Even with no magnetic field,
cooling (as measured by $\langle S\rangle$) is delayed 10\,ms by
shock heating; once the disk has settled, it commences cooling.  If
a seed field is introduced, $\langle S\rangle$ increases with time. 
The slope for the first 10\,ms is higher and quite seed
field-strength dependent and should perhaps be considered a transient
as the field saturates, while subsequent heating is slower and
less sensitive to seed field strength.

\begin{figure}
  \includegraphics[width=\linewidth]{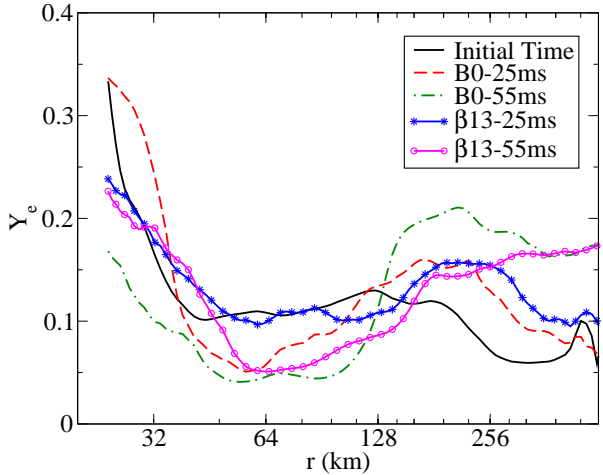}
  \caption[ye-radial]{
Vertically and azimuthally averaged electron fraction profiles
at the initial time, $t = 25\,\mathrm{ms}$, and $t = 55\,\mathrm{ms}$.
$Y_e$ decreases at densest regions and increases at low-density
  outer radii regions in nomnagnetized case.
In the magnetized case, $Y_e$ starts decreasing in the high-density
  regions only at late times.
  }
  \label{fig:Ye-radial}
\end{figure}

With no magnetic field, $\langle Y_e\rangle$ decreases monotonically,
continuing the behavior seen in our earlier
simulations~\cite{Foucart:2014nda}, while magnetized runs show a leveling
off and slight increase.  Siegel and Metzger's 3D magnetized disk
simulations also find that the inner disk remains neutron
rich~\cite{Siegel2017}.  Radial profiles of $Y_e$, displayed in
Fig.~\ref{fig:Ye-radial}, show that the magnetized disk has higher
$Y_e$ mostly in a region around radius $r\approx 40$\,km.  This can
be understood from the equilibrium electron fraction $Y_{e,eq}$.  In
this region, the magnetized disk has lower $\rho_0$ and higher $T$. 
As shown in Fig.18 of~\cite{Foucart:2014nda}, $Y_{e,eq}$ increases
with $T$ and decreases with $\rho_0$, so the higher $Y_e$ is
consistent with $Y_{e,eq}$.  The outer regions of the disk, on the
other hand, are too cool for $Y_e$ to equilibrate on the simulated
timescale.

\begin{figure}
  \includegraphics[width=\linewidth]{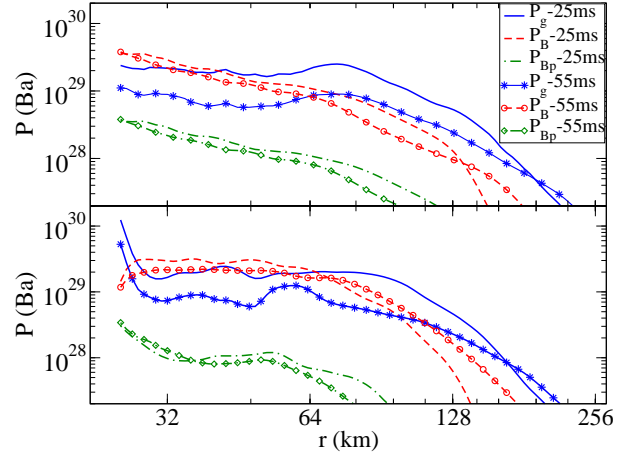}
  \caption[b2-radial]{
Vertically and azimuthally averaged radial profiles
of gas pressure, total and poloidal component magnetic pressures
at $t = 25\,\mathrm{ms}$, and $t = 55\,\mathrm{ms}$
for $\beta13$ (top panel) and $\beta36$ (bottom panel) magnetized cases.
The toroidal component becomes dominant after the magnetic field saturates
in all the magnetized cases. 
  }
  \label{fig:b2-radial}
\end{figure}

Figure~\ref{fig:b2-radial} shows gas pressure, total and poloidal magnetic pressures 
versus distance from the black hole at two times in the two
configurations $\beta13$ (top panel) and $\beta36$ (bottem panlel). The
toroidal field quicky grows to be the dominant component, contributing
about $90\%$ of the total magnetic energy at the late time evolution
for our stronger field case (as seen in either puncture or multipatch
runs).  This figure also shows that the total field pressure
exceeds the gas pressure in the inner regions.   Because of our rather
large seed fields, the field can only grow one to two orders of
magnitude before reaching overall equipartition with the internal
energy.  Strong toroidal fields can suppress the MRI, especially at
low wavenumbers~\cite{Pessah:2004de}, and this suppression may
take place in some regions of our disk.

%\begin{figure}
%  \includegraphics[width=\linewidth]{figures/rho-radial-L2.pdf}
%  \caption[rho-radial]{
%Vertically and azimuthally averaged density profiles
%at three different time snapshots:
%the initial time ($t=15\,\mathrm{ms}$ after merger),
%$t=25\,\mathrm{ms}$ where the magnetic field saturates, and $t=55\,\mathrm{ms}$ 
%which represents the late time evolution.
%The nonmagnetized case evolves to higher density at radii where the maximum density
%is located, while the magnetized case decreases in density due to the MHD-driven expansion.
%On the other hand, the average density increases near the inner edge due to the
%high accretion rate.
%  }
%  \label{fig:rho-radial}
%\end{figure}

% Neutrino emissions
\subsection{Neutrino emission and optical depth}

\begin{figure}
  \includegraphics[width=\linewidth]{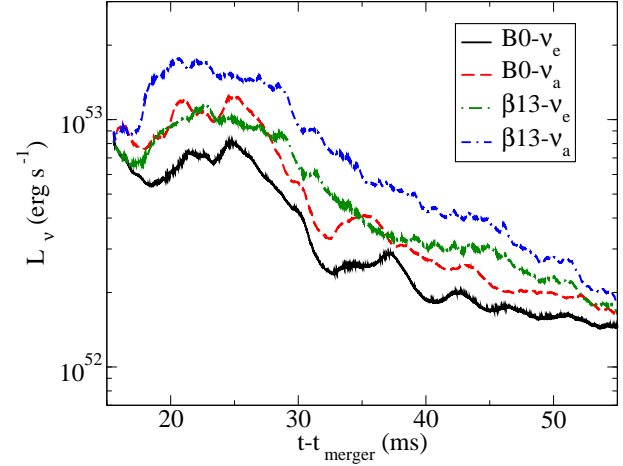}
  \caption[Lnu-type]{
   Neutrino luminosity evolution for electron-flavor neutrinos and antineutrinos 
   in the $\beta=13$ and nonmagnetized simulations.
   The magnetized run has systematically higher electron neutrino and
   antineutrino luminosities.
  }
  \label{fig:Lnu-type}
\end{figure}

\begin{figure}
  \includegraphics[width=\linewidth]{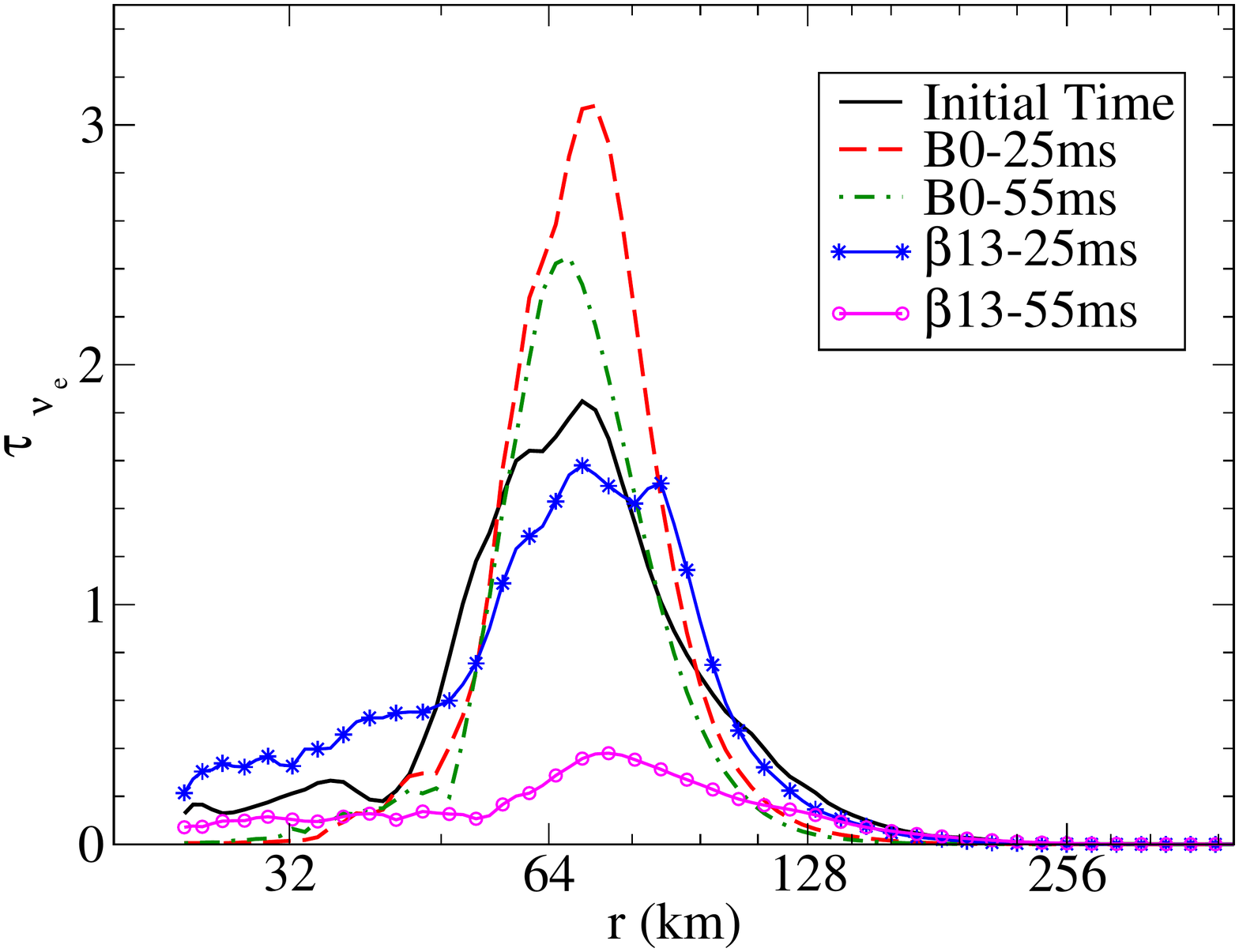}
  \caption[taux-radial]{
    The energy-averaged optical depth of electron neutrinos
    radial profiles at the initial time, $t = 25\,\mathrm{ms}$, and $t = 55\,\mathrm{ms}$.
    The nonmagnetized case becomes more opaque in the high-density
    region. This makes the neutrino cooling less efficient than in the
    magnetized case, which becomes more transparent during the evolution.
  }
  \label{fig:taux-radial}
\end{figure}

Fig~\ref{fig:Lnu-type} shows the neutrino luminosity for electron-flavor species.  The
electron antineutrino luminosity
is the strongest in both the magnetized and nonmagnetized cases. The neutrino luminosity is higher in the
magnetized case for all the species, but the changes are not as large as might have been
expected.  In all cases, the total neutrino luminosity drops from about $10^{53}\,\mathrm{erg\,s}^{-1}$ to
a few times $10^{52}\,\mathrm{erg\,s}^{-1}$ over about $30\,\mathrm{ms}$ after merger.
Radial emission profiles show that the luminosity drops
by a comparable factor throughout the high-density region; 
The drop in emission does not reflect some local effect, but
rather the global evolution of the disk: the contributions to the
luminosity are distributed smoothly throughout the high-density region.
%the drop in emission does not
%reflect some local effect--in fact, the luminosity is distributed smoothly through the
%high-density radii--but rather the global evolution of the disk.

One possible influence on $L_{\nu}$ would be a change in the neutrino optical depth.
Figure~\ref{fig:taux-radial} plots the energy-averaged optical depth of electron
neutrinos (the only neutrino flavor with optical depth sometimes greater than unity).
The nonmagnetized disk maintains an optical depth of a few, while spreading of the
magnetized disk makes it optically thin. Our disk has too low density to show
the optically thin to optically thick transition from the inner radii to the outer radii seen in
some alpha disk studies~\cite{Lee:2005se,janiuk:07}.  On the other hand, the
total neutrino luminosity, and the fact that electron anti-neutrino emission is
brightest, are consistent with the literature for $\alpha\sim 0.01-0.1$ disks~\cite{Lee:2005se,Fernandez2013}.
% the equivalent alpha I have calculated for our magnetized case at t=20,000 r=100 is about 0.3
% Isn't it too big?

%As the density rises in the inner regions of the disk, the electron fraction Ye initially drops as neutronization becomes more important, with the equilibrium composition being determined by the equality of electron and positron captures onto free neutrons and protons (see equation (6) and the discussion preceding it). The lowest value is reached at r ≃ r∗, where Ye ≃ 0.03. Thereafter it rises again, reaching Ye ≃ 0.1 close to the horizon. Thus flows that are optically thin everywhere will reach a higher degree of neutronization close to the black hole than those which experience a transition to the opaque regime. The numerical values for the electron fraction at the transition radius and at the inner boundary are largely insensitive to α, as long as the transition does occur. Lee et al 2005. p.6

Based on the accretion rate observations for the strongly magnetized case we estimate the equivalent 
$\alpha$ parameter for our disk is $\sim 0.3$.
This $\alpha$ is measured from the accretion rate, which in fact, includes the
angular momentum transport due to the magnetically-driven winds and the MRI turbulence. 
The accretion efficiency $L_{\nu}/\dot{M}c^2$ for the stronger magnetic field case is $ \geq 15\%$.
This efficiency is a few percent higher than the optically thin NDAF $\alpha$ disk models 
($\alpha \approx 0.01-0.1$)
with high spin black holes $a \geq 0.9$ as reported by Shibata et al.~(2007)~\cite{2007PThPh.118..257S} 

% Thermal evolution 
\subsection{Thermal evolution}

The transport mechanism in an accretion disk affects the luminosity in
two ways.  By heating the disk, it tends to increase the luminosity. 
By spreading the disk to larger radii and lower densities and
by facilitating higher accretion rates onto the black hole, it tends to decrease the luminosity.
For a thin alpha disk, $\tau_{\rm thermal}\ll \tau_{\rm viscous}$, so the former effect should
initially dominate, but our disk is quite thick ($H/r\approx 0.3$), so the timescales on
which these effects operate are not well separated.  To understand the actual disk evolution,
we must quantify the major heating and cooling effects.

\begin{figure}
  \includegraphics[width=\linewidth]{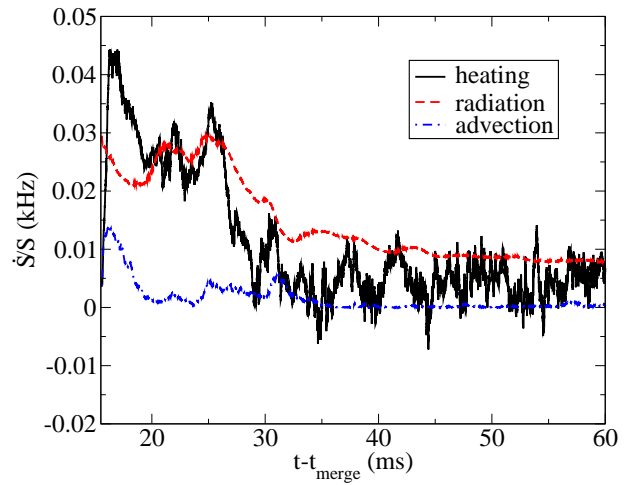}
  \caption[entropy-budget-B0]{ The evolution of the total heating rate 
    $\dot{S}^{+}$,  neutrino
    cooling rate $\dot{S}^{-}{}_{\nu}$ and advection cooling rate
    $\dot{S}^{-}{}_{Adv}$ ratios to the total entropy for the
    nonmagnetized case.  Heating and neutrino cooling rates drop
    significantly around $t = 30\,\mathrm{ms}$.  The advection cooling
    is almost zero at the end of the simulation.  }
  \label{fig:dsdt-beta0}
\end{figure}

\begin{figure}
  \includegraphics[width=\linewidth]{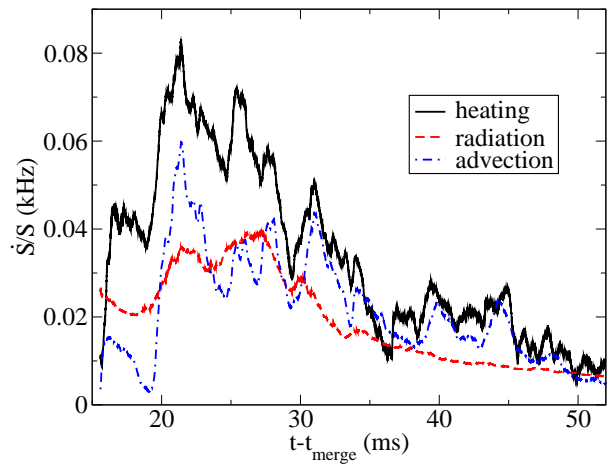}
  \caption[entropy-budget-B36]{ The evolution of the total heating rate 
    $\dot{S}^{+}$, neutrino
    cooling rate $\dot{S}^{-}{}_{\nu}$ and advection cooling rate
    $\dot{S}^{-}{}_{Adv}$ ratios to the total entropy for the case with
    weaker seed field ($\beta=36$).  
    Like in the nonmagnetized case, the total heating and
    neutrino cooling rates decrease significantly comparing with the
    early time.  The advection cooling rate is considerably higher
    due to the MHD effects.  }
  \label{fig:dsdt-beta80}
\end{figure}

\begin{figure}
  \includegraphics[width=\linewidth]{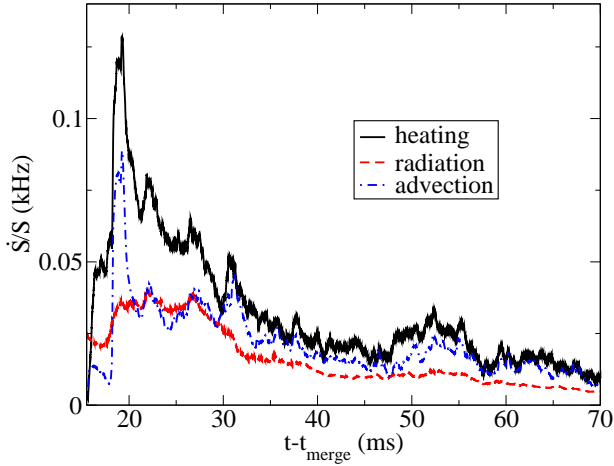}
  \caption[entropy budget-B13]{
    The evolution of the total heating rate $\dot{S}^{+}$,
    neutrino cooling rate $\dot{S}^{-}{}_{\nu}$
    and advection cooling rate $\dot{S}^{-}{}_{Adv}$ ratios to the total entropy
    for the standard, stronger seed field.
    The disk shows the same qualitative thermal behavior as in the
     weaker seed field case.
  }
  \label{fig:dsdt-beta15}
\end{figure}

Figures~\ref{fig:dsdt-beta0},~\ref{fig:dsdt-beta80}, and~\ref{fig:dsdt-beta15} show the major
entropy sources and sinks for different levels of initial magnetization.  From the energy and
lepton number source terms provided by the leakage code, a radiative entropy sink term $\dot{S}^{-}{}_{\nu}$ can
be computed (see Appendix for details).  Cooling from advection into the black hole $\dot{S}^{-}{}_{Adv}$ is
straightforwardly measured by monitoring entropy flux at the inner boundary.  Adiabatic
expansion and nuclear reactions (in nuclear statistical equilibrium) 
do not affect entropy, while shocks, reconnection, and turbulent
dissipation should only heat.  Thus, the total heating rate $\dot{S}^{+}$ should be
\begin{equation}
  \dot{S}^{+} = \dot{S} + \dot{S}^{-}{}_{\nu} + \dot{S}^{-}{}_{Adv},
\end{equation}
where $\dot{S}$ is the time derivative of the total entropy.
Unfortunately, it is difficult to separate the various possible heating sources, as they will all
appear in the code via the stabilizing dissipation terms in our shock-capturing MHD scheme.  We
normalize each source term by the instantaneous total entropy of the disk, giving the source terms
the quality of inverse timescales.

The entropy budget plots 
Figs.~\ref{fig:dsdt-beta0},~\ref{fig:dsdt-beta80}, and~\ref{fig:dsdt-beta15}
tell a clear story.
At early times, there is strong heating in all
cases from shocks as the disk, still nonlinearly perturbed from equilibrium, pulsates and
axisymmetrizes.  This heating ceases about 30\,ms after merger as the disk settles.  It is
especially clear in the nonmagnetized case (Fig.~\ref{fig:dsdt-beta0}) that this happens
before the neutrino luminosity drops.  The neutrino luminosity drops quickly thereafter,
on a fraction of the initial thermal timescale, as radiation cools the disk enough to
decrease itself.  This rapid cooling stops when the thermal timescale has increased to
about $100$\,ms.

It is worth mentioning that the unmagnetized simulations show this
initial strong heating regardless of the grid and methods used.
Indeed, it is seen even in the  original simulation presented
in~\cite{Foucart:2014nda}.  The exact amount of early-time heating
does vary noticeably from one method to another.  It is greatest for
the multipatch runs, perhaps because of numerical perturbations caused
by switching to a radically different grid
(Fig.~\ref{fig:S-Cases-Methods}).

For the nonmagnetized case, the final state is neutrino
cooling-dominated.  Accretion has nearly stopped, and advective
cooling is negligible.  The heating rate is significantly lower than
the neutrino cooling rate, although the average of the former is still
around a third of the latter.  Note that the heating rate does
occasionally become negative, presumably a sign of numerical error in
the difficult-to-follow thermal evolution of the gas as it becomes
ever more supersonic.  This negative heating could be removed by a
stricter lower limit on the entropy (see the Appendix for details),
but this would bias numerical error toward heating, which might have
an undesirable cummulative effect.

For magnetized cases, the heating rate remains above the neutrino
cooling rate.  However, this effect is largely cancelled by the strong
advective cooling that takes place as hot material accretes into the
black hole.  Although the component entropy sources and sinks are
larger than in the nonmagnetized case, the thermal timescale in these
cases also increases to $\sim$$100\,\mathrm{ms}$.  At late times, the
neutrino luminosity decreases slightly faster in the most highly
magnetized case, although in all cases a luminosity of around
$10^{52}$ erg s${}^{-1}$ will be maintained till the end of the evolution.

\begin{figure}
  \includegraphics[width=\linewidth]{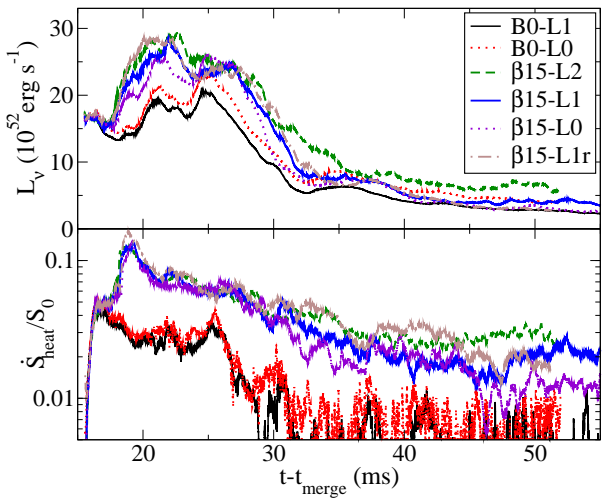}
  \caption[entropy budget-B13]{
    Convergence study of the total neutrino luminosity and heating rate for nonmagnetized and 
    strong seed field multipatch runs.
    The results are qualitatively convergent in both cases.
  }
  \label{fig:dsdt-convergence}
\end{figure}

In Figure~\ref{fig:dsdt-convergence}, we plot late-time convergence
for representative global quantities.  Convergence at these times is
difficult to achieve because it requires resolving not only the
fastest-growing MRI wavelength but also sufficient inertial range that
the average transport effects of turbulence are accurately
captured.  In the highest-resolution study to date of a BHNS
post-merger system, Kiuchi et al.~\cite{Kiuchi:2015qua} were unable to
demonstrate convergence even with grid spacing a few times smaller
than we can afford.  Thus, it is not surprising that we also obtain no
better than qualitative convergence, i.e. the overall behavior is
similar at all resolutions.  Like Kiuchi et al., we see a tendency
toward more vigorous turbulent heating at higher resolutions.

\subsection{Comparison with previous studies}

Magnetized black hole-neutron star mergers have been carried out by
other groups~\cite{Chawla:2010sw,2012PhRvD..85f4029E,Paschalidis2014,Kiuchi:2015qua}.  In particular, Etienne~{\it et
  al.}~\cite{2013arXiv1303.0837E} have also inserted a poloidal field into
a post-merger BHNS disk.  The highest-resolution MHD BHNS merger
simulation is that of Kiuchi~{\it et al}~\cite{Kiuchi:2015qua}.  These
simulations, like ours, have more realistic initial disk profiles than
analytic tori would provide.  However, there are important differences
in our treatment of the thermal evolution of the disk.  The
above-mentioned studies, since they did not employ a finite-temperature
nuclear equation of state, did not include neutrino cooling, which is
present in our simulations, so their disks were presumably too hot. 
On the other hand, by inserting a seed field only when it was safe to
apply the Cowling approximation, our disk had cooling for 15\,ms
without one of the major heating sources, so our disk is likely
over-cooled.  Furthermore, the convergence studies in~\cite{Kiuchi:2015qua}
find that heating in the inner disk is very resolution dependent,
with insufficient resolution leading to underestimates of thermally
driven outflows in particular.  A truly realistic thermal state of the
disk would presumably be somewhere in between these extremes.  (Of course,
a truly realistic treatment would also require neutrino transport,
not just a leakage approximation.)

Both the Etienne~{\it et al} poloidal seed study and the highest-resolution
simulations of Kiuchi~{\it et al}~\cite{Kiuchi:2015qua} find sustained
Blandford-Znajek Poynting flux polar jets.  Our MHD disk evolutions
do produce some unbound outflow over the simulation period
($M_{\rm ub}\sim 10^{-5}M_{\odot}$, average
$\dot M_{\rm ub}\sim 10^{-3}M_{\odot}$~s$^{-1}$) and a magnetically-dominated
polar region, but the polar field does not organize itself into a radial
Blandford-Znajek like structure.  The above-mentioned differences in
thermal treatment in our simulation and resolution effects may play a
role here.  However, we also find that the presence or absence of strong
winds and Poynting flux outflows is sensitive to the choice of seed field. 
When we evolve with a less confined initial field, we do see stronger
matter outflow and polar magnetic flux.  One might worry that these effects
could somehow be suppressed by too strong a seed field, but simulations of neutrino-cooled
disks with analytic initial conditions all find that stronger fields (as high as $\langle \beta \rangle \sim 5$)
yield stronger winds and stronger Blandford-Znajek luminosity~\cite{Janiuk:2013lna}.
Even these analytic disks with weaker initial magnetic field ($\langle \beta \rangle \sim 200$)
find strong unbounded outflows
after evolving the magnetized disk long enough~\cite{Siegel2017}, and this
might also turn out to be the case for our more confined B-field evolutions
if evolved longer.

\section{Conclusions}

We have carried out simulations of a BHNS post-merger system with a realistic initial state
provided by a numerical relativity merger simulation
, including both neutrino emission effects and
magnetic field evolution.  
The initial magnetic field is applied as large poloidal loops confined in the post-merger disk.
Because our simulations include the major heating and cooling sources,
we can study the %balance 
contribution
of each thermal driving process as the disk settles toward thermal
equilibirum.  Without a magnetic field, there is no such thermal equilibrium, so after an initial
phase of shock heating, the disk enters a phase of long-term cooling by neutrinos.  With
a strong seed magnetic field, the final state after several initial thermal timescales
is a rough balance between MHD-related heating and advective cooling, with neutrino cooling being
a secondary effect, driving the entropy down over longer timescales. 
This is roughly consistent
with the long term evolution of two dimensional neutrino cooled $\alpha$-viscosity disks 
reported by Fernandez et al.~\cite{Fernandez:2014}, where neutrino cooling is only important at early times. 
%Their alpha parameters are 0.02 and 0.05 
In both magnetized and nonmagnetized cases, the
main reason for settling is not a precise achievement of equilibrium, but an increase in
the thermal timescale (from $\sim 10$\,ms to $\sim 100$\,ms) as the initially-high neutrino
luminosity drops.

The considered magnetized 3D BHNS post-merger configuration provided 
the opportunity to test multiple methods for evolving the
relativistic MHD equations.  These show reassuring consistency over the first $\approx 20$\,ms,
but realistic long-term evolution requires careful treatment of the energy variable, especially
in how one handles the problematic recovery of primitive variables.  The multipatch methods
employed in some of our simulations can easily be applied to more general grid
configurations~\cite{Pazos:2009vb}.

The initial study of magnetized 3D BHNS post-merger disk evolution presented in this paper
is limited in many ways.  Only one BHNS system and one magnetic seed
field geometry were used.  Neutrino effects might be different for an opaque disk (e.g.,~\cite{Deaton2013}),
and magnetohydrodynamic effects are known to be seed field-dependent~\cite{Beckwith:2007sr}.
Our leakage scheme neglects
neutrino absorption, which could smooth temperature profiles and launch winds.  Existing neutrino
transport codes (e.g.,~\cite{FoucartM1:2015}) can in the future
be used to capture these effects. Finally, it would
be interesting to carry out a similar study on NSNS post-merger systems.

\acknowledgments
The authors thank Zachariah Etienne, Scott Noble, Vasileios Paschalidis, 
Jean-Pierre De Villiers, John Hawley, 
and Hotaka Shiokawa, for helpful discussions and advice over the course of
this project.  
M.D. acknowledges support through NSF Grant PHY-1402916.
F.H. acknowledges support from the Navajbai Ratan Tata Trust at IUCAA, India.
F.F. acknowledges support from Einstein Postdoctoral Fellowship grant PF4-150122, 
awarded by the Chandra X-ray Center, which is operated by the Smithsonian Astrophysical 
Observatory for NASA under contract NAS8-03060.
H.P. gratefully acknowledges support from the NSERC Canada. 
L.K. acknowledges support from NSF grants PHY-1306125 and AST-1333129
at Cornell, while the authors at Caltech acknowledge support from NSF Grants
PHY-1404569, AST-1333520, NSF-1440083, and NSF CAREER Award PHY-1151197.
Authors at both Cornell and Caltech also thank the Sherman Fairchild
Foundation for their support.  Computations were performed on the
Caltech compute clusters \emph{Zwicky} and \emph{Wheeler}, funded by
NSF MRI award No.\ PHY-0960291 and the Sherman Fairchild Foundation. 
Computations were also performed on the SDSC cluster \emph{Comet}
under NSF XSEDE allocation TG-PHY990007N.

\appendix

\section{Numerical improvements}
\label{sec:numerical-methods}

\subsection{Formulation}
The fundamental equations to be evolved are the same as in our earlier MHD
work~\cite{Muhlberger2014}.  We write the metric
\begin{equation}
ds^2 = -\alpha^2dt^2 + \gamma_{ij}(dx^i + \beta^idt)(dx^j + \beta^jdt).
\end{equation}
The fluid at each grid point is described by its set of ``primitive variables'':
baryonic density $\rho_0$, temperature $T$, electron fraction $Y_e$, and spatial components of the covariant
4-velocity $u_i$.  From $\rho_0$, $T$, and $Y_e$, the equation of state supplies
the gas pressure $P$, specific enthalpy $h$, and sound speed $c_s$.  From $u\cdot u=-1$, we
know the Lorentz factor $W=\alpha u^t$.  The stress tensor is
\begin{equation}
T_{ab} = \rho_0 h u_a u_b + P g_{ab} + F_{ac}{F_b}^c - \frac{1}{4}F^{cd}F_{cd} g_{ab}\ ,
\end{equation}
where $F_{ab}$ is the Faraday tensor.  We assume a perfectly conducting fluid,
$F^{ab}u_b = 0$, which fixes the electric field. 
The variables
actually evolved (aside from the magnetic field, whose evolution is described below)
are the conservative variables:  a density variable $\rho=\sqrt{\gamma}W\rho_0$,
the proton density $\rho Y_e$,  an energy density variable $\tau=\sqrt{\gamma}\alpha^2 T^{00} - \rho$,
and a momentum density variable $S_i=\sqrt{\gamma}\alpha T^0{}_i$.
 In the above,
%$T^{\alpha\beta}$ is
%the stress tensor, $\alpha$ the lapse, and
$\gamma$ is the determinant of the spatial metric.
We evolve using an HLLE approximate Riemann solver~\cite{HLL}.  Conservative formulations
have the advantage that numerical dissipation in shock or turbulent subscale structures is
automatically conservative.  They have the disadvantage of not evolving a separate variable
for the internal energy or entropy.  Such information must be recovered by root finding from
the conservative variables after each timestep, which can be expensive and (especially if
kinetic energy dominates over internal energy in $\tau$) inaccurate.

The magnetic field can be described via the components of its 2-form $\tilde B^i$ or its
vector field $B^i$, related as $\tilde B^i = \sqrt{\gamma} B^i$.  In a conducting medium,
field lines advect with the fluid:  $\partial_t \tilde{\bf B} = -\pounds_v\tilde{\bf B} = -d(v\cdot\tilde{\bf B})$.  Since
$d\tilde B=0$, we can alternatively evolve the vector potential 1-form ${\bf A}$, where $\tilde{\bf B}=d{\bf A}$.
A vector potential evolution will automatically satisfy $d\tilde B=0$ but will require specifying
a gauge.

Our Cartesian grid simulations suppress monopoles via a constrained transport scheme, which requires
staggering $\tilde B^i$ or $A_i$ between gridpoints.  For the multipatch simulations described below,
this would be very inconvenient because the patch coordinate transformations would have to
account for each component of the field being at a different location, so we instead code two well-known methods that control $d\tilde B$
while keeping all variables centered at the same gridpoints.  The first is a centered vector
potential method, implemented as in~\cite{Giacomazzo:2010bx}.  We find that the generalized Lorentz gauge, introduced
in~\cite{Farris:2012ux}, provides the best stability.  The evolution for $A_i$ and the scalar potential $\Phi$ are
given by
\begin{eqnarray}
  \label{AEq}
  \partial_t A_i &=& \epsilon_{ijk}v^jB^k - (\alpha\Phi-\beta^jA_j)_{,i},
  \\
  \label{PhiEq}
  \partial_t (\sqrt{\gamma}\Phi) &=& -\partial_j(\alpha\sqrt{\gamma}A^j-\sqrt{\gamma}\beta^j\Phi)
  - \xi \alpha\sqrt{\gamma}\Phi\  ,
\end{eqnarray}
where $\xi$ is a specifiable constant of order the mass of the system.  Lorentz-type gauges
lead to luminal characteristic speeds, but fortunately
the speeds used in the HLLE fluxes used in the evolution of $A_i$ (see~\cite{Giacomazzo:2010bx}) can still be
set to the physical, MHD wave maximum speed.  The signal speeds for HLLE fluxes in the $\Phi$ evolution,
on the other hand, are set to the null $-\beta^i\pm\alpha\gamma^{ii}$.

The second magnetic evolution scheme is a covariant hyperbolic divergence cleaning
method~\cite{Liebling:2010bn,Penner2011,moesta:14a}, in which
an auxiliary evolution variable $\Psi$ is introduced to damp monopoles.  The Maxwell
equation $d{\bf F}=0$ is replaced by $\star d{\bf F} = {\bf g}\cdot d\Psi - \lambda\Psi{\bf t}$,
where ${\bf g}$ is the 4-metric, ${\bf F}$ the Faraday tensor, ${\bf t}$ the unit time vector,
and $\lambda$ a specifiable damping constant.  In components
\begin{eqnarray}
  \label{BfluxEq}
  \partial_t \tilde B^i &=& \partial_i(v^j\tilde B^i - v^i\tilde B^j) + \alpha\sqrt{\gamma}\gamma^{ij}\Psi_{,j}
  + \tilde B^j{}_{,j}\beta^i,
  \\
  \label{PsiEq}
  \partial_t \Psi &=& \beta^i\Psi_{,i} - \alpha\gamma^{-1/2}\tilde B^j{}_{,j} - \lambda \Psi,
\end{eqnarray}
where we set $\lambda=1.4$.  Eq.~\ref{BfluxEq} is in conservative form and can be evolved using
our usual HLLE scheme, while Eq.~\ref{PsiEq} is evolved via straightforward second-order centered
finite differencing.

Both of these methods require added numerical dissipation.  Thus, we add Kreiss-Oliger dissipation
to the magnetic evolution equations.  For multipatch simulations, this step is done while time
derivatives are being computed
in the local patch coordinate system of evolution variable components in these coordinates.
\begin{equation}
  \partial_t X = \cdots - \Sigma_i\Delta x_i^3 D_{2i}(FD_{2i}X)\ .
\end{equation}
$X$ is $(\tilde B^i, \Psi)$ for divergence cleaning and $(A_i, \Phi)$ for the vector potential
method.  $D_{2i}$ is a second-derivative operator, and $\Delta x_i$ is the grid spacing in the $i$-th
direction, both computed in local patch coordinates.  $F$ is a function of space, which vanishes
on boundary points but may be otherwise chosen according to the problem~\cite{Mattsson:2004}.

\subsection{Cubed-sphere Multipatch Grids}
Several groups have already implemented dynamics on spherical surfaces~\cite{Sadourny1972,Ronchi1996},
3D hydrodynamics~\cite{Zink:2007xn,2011PhRvD..83d3007K,korobkin:12,reisswig:13a},
3D MHD~\cite{Koldoba:2002kx,Fragile:2008ca,Romanova:2012},
and Einstein's equations~\cite{Thornburg:2004dv,Schnetter:2006pg,Reisswig:2006nt,Pazos:2009vb,Pollney:2009yz}
with multipatch methods and cubed-sphere-like grids.  The
basic idea is to divide the computational domain into patches, each of
which has its own local coordinate system in which it is a uniform Cartesian
mesh.  In the global coordinate system, each patch is distorted, and
six distorted cubes can be fit together to fill a volume with
spherical inner and outer boundaries.  Time derivative calculations
for timesteps are computed within the local patch coordinates and
then transformed to the global coordinate system.  Multipatch methods
easily generalize to any combination of distorted cubes.  For example,
the central hole can be filled with a cube (as done in a test problem below), or
the cubed-sphere could be surrounded by non-distorted cubes.

This method can be contrasted with other popular ways of evolving grids
around black holes.  One is the use of spherical-polar coordinate grids.
The second is the use of Cartesian grids, with removal of the black hole
interior accomplished either by excising all gridpoints within a
spherical region (leading to an irregular-shaped ``legosphere'') or
by removing the interior via a radial coordinate transformation (``puncture'')~\cite{Etienne2010}.
All previous SpEC black hole-neutron star simulations use Cartesian
grids with legosphere excision.  We have been unable to find a stable
implementation of this method for magnetized flows into a black hole.
This is not surprising, since Cartesian grid faces even inside the
horizon will have characteristic fields flowing into the grid, making
the evolution ill-posed without boundary conditions providing information
about the excised interior.  Both spherical-polar and multipatch grids
can naturally excise spherical regions (which can be distorted by coordinate
transformations to fit the horizon shape as needed) and have no
incoming characteristics if placed inside the apparent horizon (and outside the
Cauchy horizon) of a stationary black hole.  Multipatch methods have an advantage over spherical-polar
grids that they do not suffer from coordinate singularities and grid pileup
near the poles, which can be an issue for high-resolution spherical-polar
simulations~\cite{Shiokawa:2011ih}.  Spherical-polar grids, on the other hand, have two
advantages.  First, for nearly axisymmetric systems, one can have much
lower resolution in longitude than in latitude, a freedom not present
in multipatch grids.  Second, communication between patches in multipatch
grids is by ghost zone overlaps.  Ghost zone gridpoints will not match
gridpoints on the overlapping live patch, so they must be filled by
interpolation.  This introduces a new source of error which will
generally not exactly respect conservation laws and may create magnetic monopoles,
although it should converge away with resolution.  Which method is
best most likely depends on the problem.

Since our conservative evolution equations are generally covariant,
it is straightforward to evolve them in the local patch coordinates,
shifting to global coordinates for ghost zone synchronization.  For
our WENO5 reconstruction method, we need three ghost zone layers
on patch interior boundaries.  Because of our methods of ``fixing''
problematic points described below, synchronizing variables
is not quite the same as just synchronizing their time derivatives,
and we find the former to be needed for stability. 
For the divergence cleaning method,
any monopoles generated by interpolation of $\tilde B^i$ in ghost
zones are damped (by design of Eqs~\ref{BfluxEq} and~\ref{PsiEq})
and remain small.
For the vector potential method, we synchronize $\tilde B^i$
computed from $A_i$ on patch faces, where information is
lacking on one side to compute the curl.  It is crucial here
to synchronize only the outermost layer of points, not the
full 3-layer ghost zone region, because the latter will
introduce monopoles in the ghost zones and lead rapidly
to an instability there.

\begin{figure}
  \includegraphics[width=1.\columnwidth]{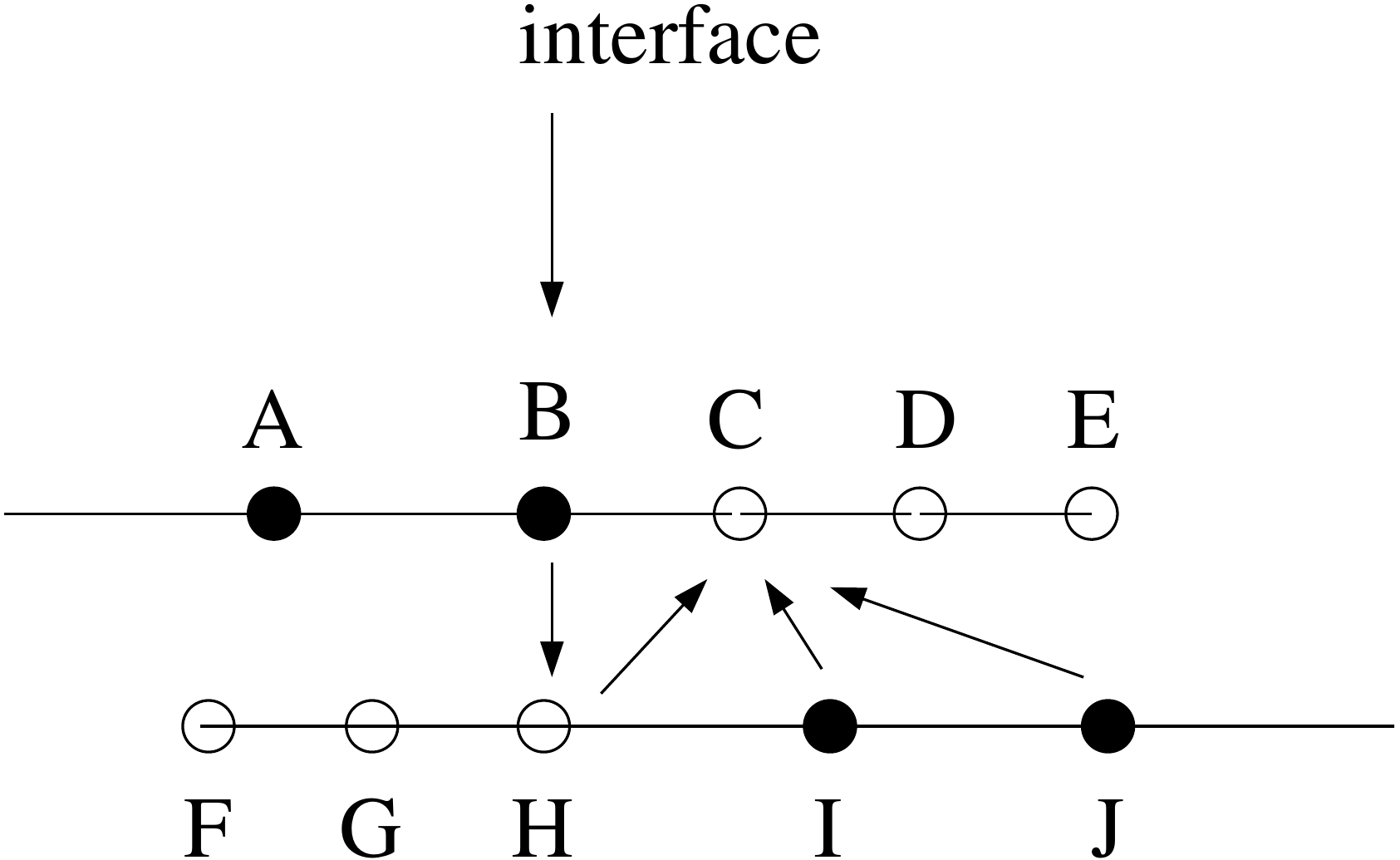}
  \caption{An illustration of the synchronization of
    ghost zone regions at internal patch boundaries.
    As is standard practice with uniform grids, one
    grid is extended the full ghost zone width (three
    points, in our case) beyond the interface, while
    the other grid extends two points.  For a cubed
    spheres setup, ghost zone extensions must be chosen
    to guarantee sufficient overlaps on 3-patch edges.
    Above, open circles are ghost zone points; filled
    circles are live points.  The points B and H
    overlap and mark the interface.  First H is set
    to B (which does not require interpolation).
    Then H can be used in the interpolation to
    get C.
    }
  \label{fig:noliveoverlap}
\end{figure}

Our cubed-sphere grids are largely the same as those of
other groups.  A minor alteration in the ghost zones
is illustrated in Fig.~\ref{fig:noliveoverlap} to eliminate the presence
of overlap regions which are ``live'' for both grids
(i.e. neither is synced with respect to the other).  Our
fears that ``live overlaps'' would be dangerous have not
been borne out, but the new arrangement does seem to propagate
shocks a bit better and show less deviation in rest mass
(interpolated ghost zones do not allow strict mass conservation
in either case), although it
cannot be generalized to more general multipatch structures.

\begin{figure}
  \includegraphics[width=\linewidth]{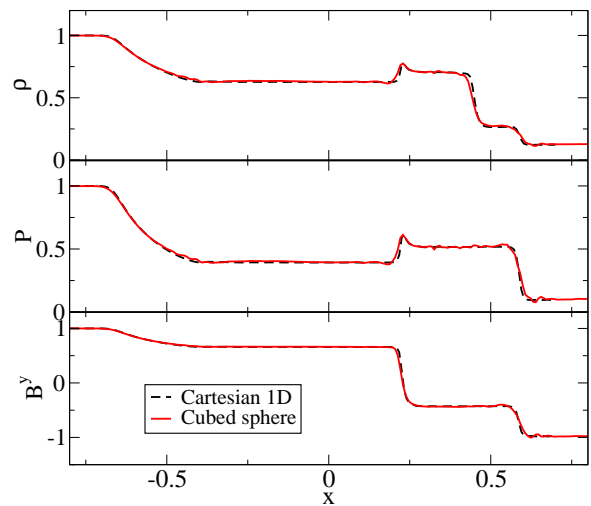}
  \caption{Magnetized Riemann problem evolved on both a cubed-sphere
    multipatch grid (with about 240 grid points across the diameter of
    the spherical computational domain), together with the results for
    the same problem evolved on a Cartesian one-dimensional grid which
    is able to utilize the planar symmetry.  The interface between the
  inner cube and outer cubed spheres is at $\pm$ 0.27.}
  \label{fig:Balsara}
\end{figure}

\begin{figure}
  \includegraphics[width=\linewidth]{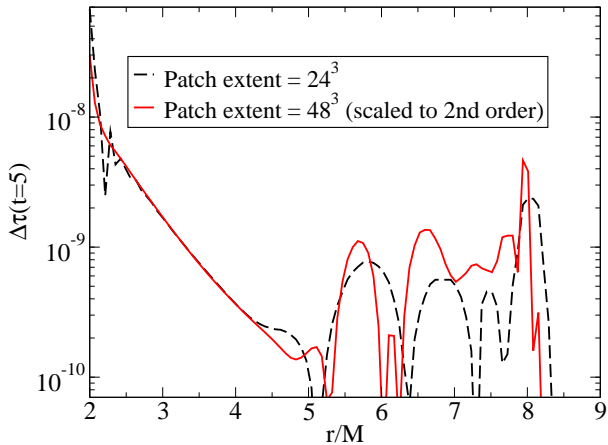}
  \caption{Convergence test for Bondi accretion with a radial magnetic
    field.  Shown here is the error in the $\tau$ conservative variable
    at $t=5M$, by which time it has settled. 
    The error plotted is the absolute change in $\tau$.  The
    relative change of $\tau$ is about $2\times 10^{-4}$ at the lower resolution.  
    Second-order convergence
    breaks down at the sonic radius at $r=8$, as expected.  The grid
    consists of 48 domains, with each of the six patches split in two
    on each of its axes.}
  \label{fig:Bondi}
\end{figure}

\begin{figure}
  \includegraphics[width=\linewidth]{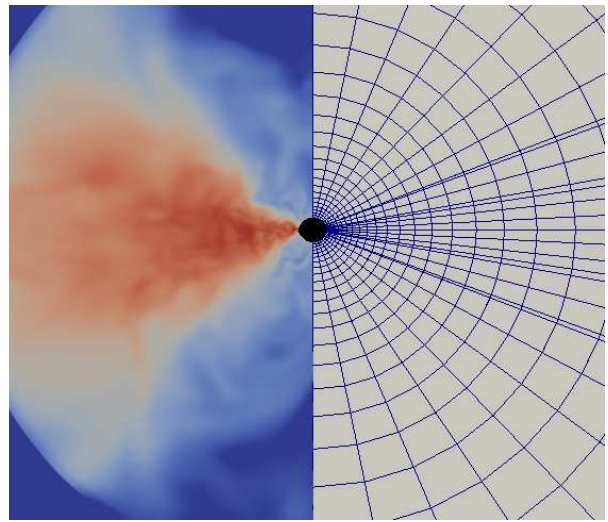}
  \caption{Meridional snapshot at $t=1600M$ of a turbulent accretion torus.  Shown
    on the left is the density (on a logarithmic scale covering the four
    decades up to the maximum).  On the right is the grid, with resolution reduced
    by about a factor of 4 for clarity.  The effects of the radial and angular maps
    are visible, as is the regularity of the poles.  The unusually close radial
    lines are nonmatching ghost zones.
  \label{fig:FMtorus}
    }
\end{figure}

Figures~\ref{fig:Balsara}, \ref{fig:Bondi}, and~\ref{fig:FMtorus} show
some standard MHD test problems applied to the multipatch MHD code.
Fig.~\ref{fig:Balsara} is the first Riemann problem
from~\cite{Balsara:2001} and~\cite{Giacomazzo2007}, containing a
left-going fast rarefaction wave, a left-going compound wave, a
contact discontinuity, a right-going slow shock and a right-going fast
rarefaction wave.  To test relativistic terms, we set lapse
$\alpha=0.5$ and shift $\beta^n=0.1$, yielding the expected slowdown
and advection.  A cubical patch is added to the center to fill the
inner hole, while the planar symmetry is imposed on the outer
boundary, setting functions in the outer points to their values at the
closest point in the interior on a line in the symmetry plane.  The
waves travel through interpolated boundaries without incident.

In Fig.~\ref{fig:Bondi}, we evolve a Bondi accretion problem (the same
as in~\cite{Muhlberger2014}) with a radial magnetic field and maximum
$\beta^{-1}$ of 2.5.  Both of these tests are performed with the
divergence cleaning code.  As in our earlier
paper~\cite{Muhlberger2014}, we find better behavior when we add
Kreiss-Oliger dissipation to all variables, with $F=0.06r^{-2}$.
Errors saturate after a few $M$ of evolution, with second-order
convergence demonstrated except at the sonic point and  the inner
boundary.

Finally, we evolve a constant angular momentum Fishbone-Moncrief torus
~\cite{Fishbone:1976}
around a rapidly spinning black hole.  We set the dimensionless spin
of the black hole to $a/M_{\rm BH}=0.938$, the angular momentum parameter
to $\ell=4.281$ ($\ell=u^t u_{\phi}$), 
average $\beta=100$, and the equation of state to a Gamma
law with $\Gamma=4/3$, making the problem very similar to a standard
scenario studied by the HARM code~\cite{mckinney:04,Shiokawa:2011ih}.
We use the same radial and angular coordinate maps as in these
studies.  Like in~\cite{Fragile:2008ca}, we find that it is necessary
to tilt the grid in order for the current sheet formed by winding of
the seed field to break in a reasonable time and initiate turbulence.
We agree with their observation that this is an artifact of symmetries
in the setup and should not be a worry for general problems.
Fig.~\ref{fig:FMtorus} shows on the left a snapshot of the density at
$t=1600M$, on the right a representation of the grid with resolution
quartered for clarity.  The actual evolution grid used 120 radial
points and 60 angular points across each of the six patches.  For this
problem, we found it advantageous to have higher dissipation in
problematic regions (low density regions and the viscinity of the
black hole) and low dissipation inside the torus, where we wanted to
resolve the MRI with modest resolution.  There we set $F=0.01$
($F=0.001$) inside the disk for $X=\tilde B^i$ ($X=A_i$), and we set
$F=0.1$ for $\rho_0/\rho_{0\rm max}<0.05$ or $r/M_{\rm BH}<3$.
Results qualitatively match the literature, with mass flow into the
horizon $<\dot{M}>\sim 1$, electromagnetic energy flux {\it out} of
the horizon $<L_{\rm EM}>\sim 10^{-2}\dot{M}$, and the generation of
unbound matter.

\subsection{Coordinate maps}
In its fluid module, SpEC assumes uniform grid spacing in the coordinates
on which the grid is defined, so nonuniformity can be acheived by
introducing coordinate transformations between these grid coordinates
and the original, ``physical'' coordinates.  All simulations use
logarithmic radial maps [i.e. uniform spacing in $\log(r)$], concentrating
grid near the black hole.  For Cartesian simulations, this naturally
introduces a puncture, set at the desired excision radius $r_{\rm EX}$.
It leads to enormous distortions on the edges of the cubical grid, but
since we only evolve in a sphere contained by the cube, this causes no
problems.  For multipatch simulations, the exponential map preserves
the ratio between radial and transverse grid spacings; both increase
with distance from the center.

We also add maps to concentrate grid near the equator.  For multipatch
runs, we use the angular map common for MHD disk simulations
$\theta=\pi \theta' +(1-h)\sin(2 \pi \theta')/2$
~\cite{mckinney:04} with
$h=0.4$.  For Cartesian runs, this angular map unacceptably distorts
grid cells, leading to artifacts in the evolution, so we instead
use a cubic scale map on the z axis ($Z
= z - \lambda(z-R_{min})^3/R_{min}^2$ with $\lambda=-0.375$ and
$R_{min}=1.0$).  

Finally, we have carried out multipatch simulations using a radial
map (composed with the logarithmic map) to concentrate grid on
a ring coinciding with the high-density region.  The map has the
form
\begin{equation}
  r'-r_0 = A \arctan\left[\frac{r-r_0}{\lambda}\right] + B (r-r_0) + C (r-r_0)^2\ ,
\end{equation}
where $r$ is the grid radius, $r'$ the physical radius, $\lambda$ controls
the width of the zoomed region, while $A$, $B$,
and $C$ are set so that $r$ and $r'$ coincide at the inner and outer
radii, and the appropriate zoom factor ($dr'/dr$) is acheived at $r'=r=r_0$. 

\subsection{Primitive variable recovery}
Sometimes, due to numerical error, the evolved conservative variables
$(\rho,\rho Y_e,\tau,S_i,\tilde B^i)$ may not correspond to any
physical $(\rho_0,T,Y_e,u_i,B^i)$.  In this case, we can ``fix'' the
conservative variables to make primitive variable recovery possible
using the prescription described in Appendix A
of~\cite{Muhlberger2014} (straightforwardly altered to take into
account the minimum of $h$ being less than one~\cite{Deaton2013}).
Unfortunately, this introduces glitches in supersonic flows such as
those in thin disks, usually seen as gridpoints at which the
temperature discontinuously jumps to the equation of state table
minimum.  Although this is initially a cooling effect, the glitches
create artificial heating.  For nonmagnetized disk simulations, this
ultimately stalls the cooling of the disk after only a small decrease
in total entropy.

We remove this problem by introducing an auxiliary entropy evolution
variable $\rho S$, where $S$ is the entropy per baryon.  The use of
entropy variables to reset problematic gridpoints and ameliorate
accuracy problems in the evolution of internal energy by conservative
codes has already been tried by other
groups~\cite{Balsara:1999,Ruffert:2001gf,Koldoba:2016}.

In the absence of subgrid-scale energy dissipation
(shocks, reconnection, turbulence), the entropy
of a fluid in nuclear statistical equilibrium evolves
by advection and neutrino emission only~\cite{1983bhwd.book.....S,Ruffert:2001gf}.
\begin{equation}
  \label{eq:entropy}
  \partial_t (\rho S) + \partial_i(\rho S v^i)
  = \frac{m_n\alpha\sqrt{\gamma}}{k_BT}[Q_{\nu} -R_{\nu}(\mu_e + \mu_p - \mu_n)]
\end{equation}
where $Q_{\nu}$ and $R_{\nu}$ are the net neutrino energy and
lepton number emission rates per volume, respectively, $m_n$
is the nucleon mass, and $\mu_X$ are chemical potentials. 
Note that, since we have excluded only heating effects,
the evolved $\rho S$ gives a lower bound on the true
entropy.

Roughly speaking, we now have two energy variables,
$\tau$ and $\rho S$, which are made to be consistent
with each other at the beginning of each timesetep.
Each step, we execute the following procedure.

\begin{enumerate}
\item Evolve $(\rho, \rho Ye, \rho S, \tau,S_i, \tilde B^i)$
  using an HLL approximate Riemann solver. 
  $\rho S$ must be evolved with a monotonic reconstructor to
  avoid new extrema.  We use a second-order monotonized centered (MC2)
  limiter~\cite{MC}.  The other variables
  can be evolved with higher-order reconstruction like WENO.
\item Compute $S$, $Y_e$, and $B^i$ from the appropriate
  divisions of the conservative variables.
\item If not, attempt to solve $(T,W^2)$ using $\tau$
  and the other conservative variables except $\rho S$
  using the \texttt{gnewton} method as implemented by the GSL Scientific
  Library~\cite{gslmanual}.
\item If a root is found, use it to compute the entropy,
  $S_{\tau}$.  If $S_{\tau}>\chi S$, accept the root.  The
  parameter $\chi\le 1$ but is otherwise freely specifiable.
  We use $\chi=0.97$.
\item If a root was not found, or if it violates the condition
  in step 4, first check to see if the point is in the force-free
  regime.  If so, use force-free recovery of $(T,W)$.  (See~\cite{Muhlberger2014}
  for details on this solver and the conditions for its use.)
\item If the point does not meet the force-free conditions,
  attempt to solve for $(T^3,W^2)$ using $\rho S$
  and the other conservative variables except $\tau$, again
  using GSL's \texttt{gnewton}.  If
  a root was found in step 4, use it as the initial guess
  for the root solve.  (Using $T^3$ instead of $T$ speeds up
  convergence in some difficult points, but probably makes
  little difference in general.)
\item If a root could not be found with multidimensional
  root finding, attempt again with $\rho S$ and other
  variables except $\tau$, this time with GSL's 1D
  \texttt{brent} root finder.  Here we regard $W$ as
  the variable, solving Eq. (A24) of~\cite{Muhlberger2014}, with
  $T$ solved via a separate 1D solve of the condition
  $S=S(\rho_0,T,Y_e)$ on each iteration.  This 1+1D
  solving is much slower but more robust than the 2D solver.
\item If this fails, attempt a 1D bracketing algorithm
  for $h\rho_0W^2$ which uses $\tau$ rather than $S$.
  (See Appendix A of~\cite{Muhlberger2014}).  If this fails, terminate
  the evolution with an error.
\item If an acceptable root was found, apply other
  ``atmosphere'' fixes to the primitive variables at low densities:  limits to
  the temperature and Lorentz factor in these regions.
\item Recompute all conservative variables from these
  final primitive variables.  $\tau$ and $\rho S$ are now
  again consistent.
\end{enumerate}

A simple sanity check on our implementation of the source terms in
Eq.~\ref{eq:entropy} is to alter the above to force the code to always
use the evolved $S$ in primitive variable recovery, in which case one
observes the disk cooling on a timescale of the total thermal energy
divided by neutrino luminosity.

In Fig.~\ref{fig:S-Cases-Methods}, we have already shown the
difference this method makes to the entropy evolution of the
nonmagnetized disk.  Significantly, all discontinuous artifacts are
gone when the new method is used.  Because the magnetized disk does
not reach such low entropies, the choice of methods makes little
difference for those simulations.
  
\bibliography{References/References}

\end{document}